\documentclass[sigconf]{acmart}

\setcopyright{none}

\renewcommand\footnotetextcopyrightpermission[1]{}
\usepackage{xspace}
\usepackage{hyperref}
\usepackage{url}
\usepackage{multirow}
\usepackage{graphicx}
\usepackage{array}
\usepackage{makecell}
\usepackage{booktabs}
\usepackage{tabularx}
\usepackage{ragged2e}
\usepackage{booktabs}
\usepackage{dashrule}
\usepackage{algorithm}
\usepackage{algpseudocode}
\usepackage[table]{xcolor}
\usepackage{pifont}
\usepackage{arydshln}
\usepackage[most]{tcolorbox}
\usetikzlibrary{shadows}
\usepackage{xcolor}
\newcommand{\ourmethod}{{HVM-GraphRAG}\xspace}
\AtBeginDocument{%
  }

\definecolor{promptBlue}{RGB}{112,124,195}

\newtcolorbox{promptbox}[1]{
  enhanced,
  colback=white,
  colframe=promptBlue,
  coltitle=white,
  colbacktitle=promptBlue,
  title=\textbf{#1},
  fonttitle=\large\bfseries,
  boxrule=2.2pt,
  titlerule=0pt,
  arc=0pt,
  outer arc=0pt,
  left=10pt,
  right=10pt,
  top=8pt,
  bottom=8pt,
  toptitle=6pt,
  bottomtitle=6pt,
  boxsep=0pt,
  sharp corners,
  drop fuzzy shadow,
}

\definecolor{relationPurple}{RGB}{126, 102, 184}

\newtcolorbox{relationpromptbox}[1]{
  enhanced,
  colback=white,
  colframe=relationPurple,
  coltitle=white,
  colbacktitle=relationPurple,
  title=\textbf{#1},
  fonttitle=\large\bfseries,
  boxrule=2.2pt,
  titlerule=0pt,
  arc=0pt,
  outer arc=0pt,
  left=10pt,
  right=10pt,
  top=8pt,
  bottom=8pt,
  toptitle=6pt,
  bottomtitle=6pt,
  boxsep=0pt,
  sharp corners,
  drop fuzzy shadow,
}

\definecolor{conflictRed}{RGB}{190, 92, 88}

\newtcolorbox{conflictpromptbox}[1]{
  enhanced,
  colback=white,
  colframe=conflictRed,
  coltitle=white,
  colbacktitle=conflictRed,
  title=\textbf{#1},
  fonttitle=\large\bfseries,
  boxrule=2.2pt,
  titlerule=0pt,
  arc=0pt,
  outer arc=0pt,
  left=10pt,
  right=10pt,
  top=8pt,
  bottom=8pt,
  toptitle=6pt,
  bottomtitle=6pt,
  boxsep=0pt,
  sharp corners,
  drop fuzzy shadow,
}

\definecolor{resolveTeal}{RGB}{72, 150, 142}

\newtcolorbox{resolvepromptbox}[1]{
  enhanced,
  colback=white,
  colframe=resolveTeal,
  coltitle=white,
  colbacktitle=resolveTeal,
  title=\textbf{#1},
  fonttitle=\large\bfseries,
  boxrule=2.2pt,
  titlerule=0pt,
  arc=0pt,
  outer arc=0pt,
  left=10pt,
  right=10pt,
  top=8pt,
  bottom=8pt,
  toptitle=6pt,
  bottomtitle=6pt,
  boxsep=0pt,
  sharp corners,
  drop fuzzy shadow,
}

\definecolor{subRed}{RGB}{180,90,78}
\definecolor{subCyan}{RGB}{35,135,145}
\definecolor{lightgreen}{RGB}{210, 245, 210}
\definecolor{lightyellow}{RGB}{255, 252, 230}
\definecolor{lightred}{RGB}{255, 232, 236}
\definecolor{lightpurple}{RGB}{238, 229, 238}

\definecolor{darkgreen}{RGB}{190, 230, 190}
\definecolor{darkyellow}{RGB}{248, 246, 210}
\definecolor{darkred}{RGB}{245, 210, 216}
\definecolor{darkpurple}{RGB}{220, 205, 220}

\setcopyright{acmlicensed}
\copyrightyear{2018}
\acmYear{2018}
\acmDOI{XXXXXXX.XXXXXXX}
\acmConference[Conference acronym 'XX]{Make sure to enter the correct
  conference title from your rights confirmation email}{June 03--05,
  2018}{Woodstock, NY}
\acmISBN{978-1-4503-XXXX-X/2018/06}

\begin{document}

\title{HVM-GraphRAG: Holistic-View Multimodal Graph Retrieval-Augmented Generation on Complex Document}

\author{Xin He}
\affiliation{%
  \institution{Jilin University}
  \city{Changchun}
  \country{China}
}
\email{hex23@mails.jlu.edu.cn}

\author{Yili Wang}
\affiliation{%
  \institution{Jilin University}
  \city{Changchun}
  \country{China}
}
\email{wangyili@jlu.edu.cn}

\author{Wenqi Fan}
\affiliation{%
  \institution{The Hong Kong Polytechnic University}
  \city{Hong Kong SAR}
  \country{China}
}
\email{wenqifan03@gmail.com}

\author{Qing Li}
\affiliation{%
  \institution{The Hong Kong Polytechnic University}
  \city{Hong Kong SAR}
  \country{China}
}
\email{csqli@comp.polyu.edu.hk}

\author{Qinggang Zhang}
\authornote{Corresponding author.}
\affiliation{%
  \institution{Jilin University}
  \city{Changchun}
  \country{China}
}
\email{qinggangzhang@jlu.edu.cn}

\author{Yi Chang}
\affiliation{%
  \institution{Jilin University}
  \city{Changchun}
  \country{China}
}
\email{yichang@jlu.edu.cn}

\author{Xin Wang}
\authornotemark[1]
\affiliation{%
  \institution{Jilin University}
  \city{Changchun}
  \country{China}
}
\email{xinwang@jlu.edu.cn}




\begin{abstract}
Question answering (QA) over complex documents requires models to retrieve and integrate evidence distributed across distant document regions and modalities.
Multimodal GraphRAG provides a promising direction by organizing document evidence with graph structures. 
However, existing methods often suffer from unreliable cross-modal evidence indexing and expensive graph traversal. 
To address these issues, we propose \ourmethod, a holistic-view multimodal GraphRAG framework on complex document. 
\ourmethod uses a holistic view to guide graph construction, thereby reducing noisy and conflicting graph updates and building reliable indices between concept-level graph nodes and supporting multimodal chunks.
During retrieval, \ourmethod searches over a compact concept-level graph and directly accesses supporting evidence through the constructed index, avoiding costly traversal over dense entity-level graphs. 
After obtaining the retrieved evidence, \ourmethod further reorganizes chunks into modality-specific groups, enabling the answering model to better integrate heterogeneous evidence.
Experiments on three datasets show that \ourmethod achieves the best answer performance in most evaluated settings while substantially improving online retrieval efficiency over representative graph-based baselines.
\end{abstract}


\maketitle

\section{Introduction}
\begin{figure}[!t]
  \includegraphics[width=1.0\linewidth]{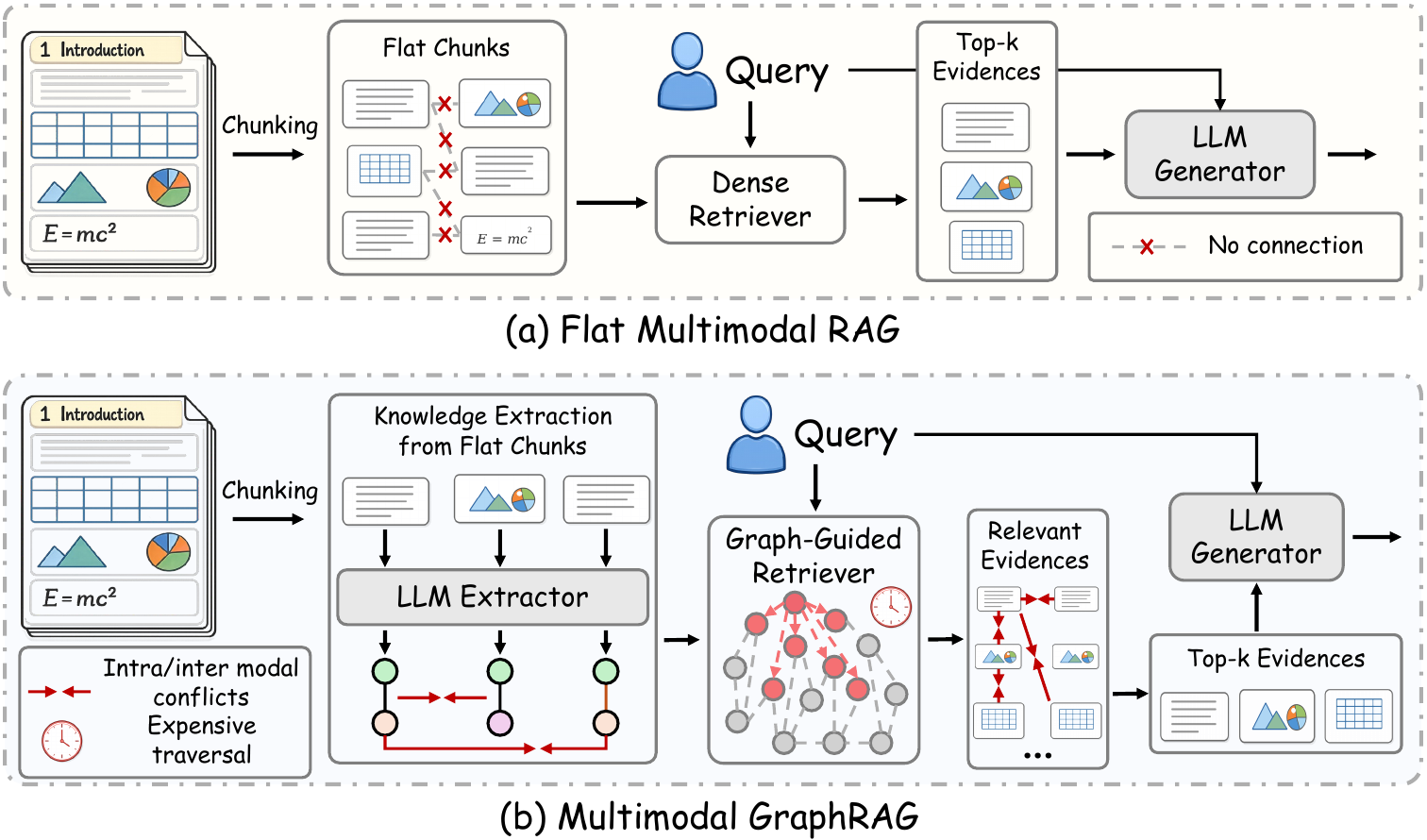}
  \vspace{-8mm}
  \caption{Flat Multimodal RAG and Multimodal GraphRAG. 
  (a) Flat Multimodal RAG loses document hierarchy and cross-modal dependencies by retrieving isolated chunks. (b) Multimodal GraphRAG introduces structural knowledge, but may suffer from unresolved conflicts and costly graph traversal.
  }
\Description{}\label{fig:flat_Multimodal_rag_and_Multimodal_graph_rag}
\vspace{-6mm}
\end{figure}

Complex document question answering (QA)~\cite{deng2025longdocurl,yu2025docthinker} is a fundamental task for accessing and synthesizing knowledge in real-world information systems. 
Its applications span scientific literature analysis~\cite{zheng2025information,tang2026ai}, financial report analysis~\cite{li2025extracting,choi2025finagentbench}, legal document understanding~\cite{zhang2025syler,sadowski2025verifiable,barron2025bridging}, and technical manual assistance~\cite{zhang2025domain,bai2025qwen3}.
Unlike short-passage QA~\cite{wu2026memgraphrag,chen2026legalgraphrag}, complex document QA requires integrating evidence distributed across distant document sections and heterogeneous modalities, including text, tables and images.
Multimodal Retrieval-Augmented Generation (Multimodal RAG) provides a general paradigm for this task~\cite{riedler2024beyond,xia2025mmed,chen2022murag}: it retrieves question-relevant evidence from multimodal document content and then generates answers grounded in the retrieved evidence.
However, most existing Multimodal RAG methods rely on flat chunk retrieval (Fig.\ref{fig:flat_Multimodal_rag_and_Multimodal_graph_rag}(a)), which treats multimodal chunks as isolated evidence units~\cite{robertson2009probabilistic,quinn2025accelerating}. 
Such isolated evidence modeling often fails to capture document-level structures, cross-modal correspondences, and semantic relations among evidence units~\cite{wu2026memgraphrag,yang2025graphusion}.

Recent Multimodal RAG studies~\cite{bu2025query,hsiao2026megarag,yuan2025mkg,wang2025bookrag} have therefore explored graph structures for organizing multimodal evidence with explicit nodes and edges.
In this paradigm, the graph provides a structured organization layer for modeling relations among multimodal information, whether the information is represented as extracted knowledge or retained as source evidence.
Existing Multimodal GraphRAG methods can be broadly grouped into two categories according to whether the graph primarily represents extracted knowledge or indexes source evidence~\cite{peng2025graph,zhang2025survey}.
The first category is \textbf{Knowledge-based Multimodal GraphRAG}~\cite{wan2025mmgraphrag,dai2026mg,yu2026can}, which constructs or leverages multimodal knowledge graphs by extracting entities, attributes, and relations from multimodal content.
The second category is \textbf{Index-based Multimodal GraphRAG}~\cite{wang2025bookrag,wu2025molorag,liu2025graph}, in which graphs serve as indexing structures over original multimodal evidence units. 
Their topology guides evidence localization, expansion, and aggregation. 
Overall, graph structures shift Multimodal RAG from independent chunk matching to relation-aware evidence modeling, which is crucial when answers require multi-hop evidence aggregation across pages, regions, and modalities.

\begin{figure}[!t]
  \includegraphics[width=0.95\linewidth]{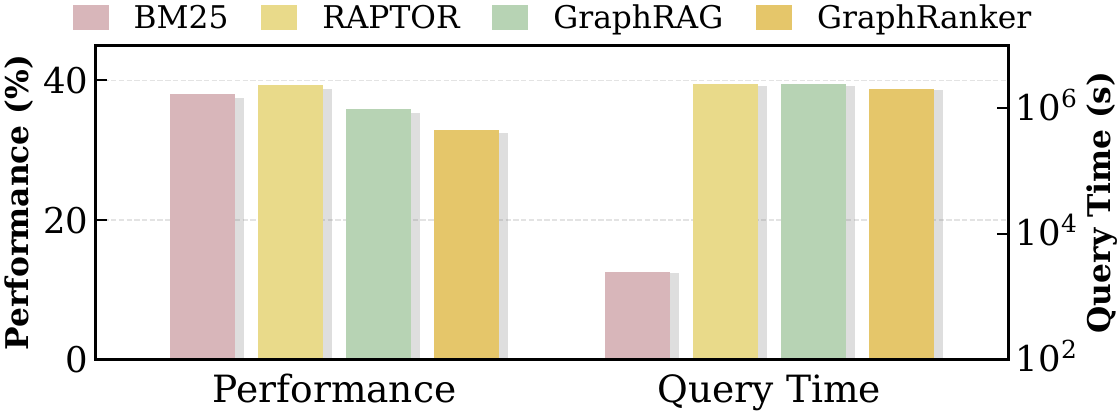}
  \vspace{-4mm}
  \caption{Comparison of answer accuracy and total online query
time on Qasper.
  }
\label{fig:performance_and_query_time_between_different_model}
\vspace{-6mm}
\end{figure}

Despite these structural advantages, existing graph-based RAG methods~\cite{sarthi2024raptor,edge2404local,gutierrez2024hipporag} do not consistently outperform the flat retrieval method in complex document QA (Fig.~\ref{fig:performance_and_query_time_between_different_model}). 
Their practical benefits are often limited by two challenges introduced during graph construction and retrieval.
The first challenge is \textbf{Unreliable Cross-Modal Evidence Indexing}~\cite{wu2026memgraphrag}. 
Existing methods often extract knowledge from multimodal chunks independently and insert it into the graph without a holistic view of the document-wide graph state (Fig.~\ref{fig:flat_Multimodal_rag_and_Multimodal_graph_rag}(b)). 
Consequently, locally extracted concepts, entities, and relations are not systematically checked against existing graph elements or evidence from other modalities. 
This may introduce noisy, redundant, or conflicting graph structures and unreliable graph-to-chunk indices, ultimately leading to suboptimal question-answering performance.
The second challenge is \textbf{Overly Expensive Graph Traversal}~\cite{hu2026remindrag,li2026scout}. 
As complex documents contain many pages, regions, and entities, the constructed graph can become large and dense. 
Retrieving evidence from such a graph often requires costly node matching, relation traversal, neighborhood expansion, and evidence aggregation (Fig.\ref{fig:flat_Multimodal_rag_and_Multimodal_graph_rag}(b)). 
This substantially increases inference latency and reduces retrieval efficiency, especially when repeated graph expansion is required for each question.

To address these challenges, we propose \textbf{\ourmethod}, a \textbf{H}olistic-\textbf{V}iew \textbf{M}ultimodal \textbf{GraphRAG} framework that uses holistic context to construct a cleaner concept graph, improving evidence-indexing reliability and retrieval efficiency for question answering over complex documents.
Specifically, \ourmethod consists of two main modules: \ding{182} The Holistic-View-Guided Graph Construction Module (HGCM) incorporates document-level structural and semantic context into graph construction, building a holistically coherent concept graph with reliable evidence indexing.
\ding{183} The Graph-Guided Holistic Retrieval Module (GHRM) retrieves query-relevant concept nodes from the compact concept graph and expands them to supporting multimodal evidence through the evidence index, reducing traversal cost for answer generation.
Together, these modules construct a cleaner concept graph and perform retrieval over a more compact concept space, thereby improving answer reliability and inference efficiency.

The contributions of this work are summarized as follows:

\begin{itemize}
    \item We identify key limitations of existing Multimodal GraphRAG methods on complex document: Unreliable Cross-Modal Evidence Indexing, and Overly Expensive Graph Traversal.

    \item We propose \textbf{\ourmethod}, a holistic-view Multimodal GraphRAG framework that improves answer accuracy and retrieval efficiency through Holistic-View Graph Construction and Graph-Guided Holistic Retrieval.

    \item We conduct comprehensive experiments on three complex document question-answering datasets, demonstrating that \ourmethod improves both answer accuracy and inference efficiency over existing baselines.
\end{itemize}

\section{Preliminary Study}
This section defines the core components of our knowledge representation, formalizes complex document QA within a two-stage GraphRAG framework, and empirically analyzes the limitations of existing Multimodal GraphRAG methods.
\subsection{Definitions}
To facilitate the presentation of our method, we briefly formalize the core components of our knowledge representation.

\textbf{Concept ($c$) and Entity ($e$)}: A \textit{concept} $c$ denotes a high-level category, while an \textit{entity} $e$ denotes a concrete document instance. 
Each entity is assigned to exactly one concept by a concept assignment function $\phi(e)$.

\textbf{Schema ($s$) and Fact ($f$)}: A \textit{schema} $s=(c_h,r,c_t)$ represents a concept-level relation between two concepts. 
A \textit{fact} $f=(e_h,r,e_t)$ instantiates a schema at the entity level.

\textbf{Chunk ($b$)}: A chunk $b=(x,m)$ denotes a basic document unit, where $x$ represents its content and $m\in\Omega$ specifies its modality. 
The modality set is defined as $\Omega=\{\text{text},\text{table},\text{image}\}$.

Detailed definitions and examples are provided in Appendix~\ref{app:definitions}.

\subsection{Problem Statement}
Complex document QA~\cite{zhao2024docmath} often requires integrating evidence distributed across chunks, pages, and modalities.
To support such evidence aggregation, Multimodal GraphRAG organizes document information into graph structures and uses them to guide evidence retrieval. 
We formulate the process as two stages: offline graph construction and online graph-guided retrieval and answering.

\textbf{Offline Graph Construction.}
Given a complex document $D$, the offline stage organizes its multimodal chunks into a graph-based representation for evidence indexing. 
Formally, the construction process produces a graph $\mathcal{G}=(\mathcal{V}, \mathcal{E})$ and an evidence index $\mathcal{I}^{*}$:
\[
(\mathcal{G}, \mathcal{I}^{*}) = \text{GraphConstructor}(D),
\]
where $\mathcal{V}$ denotes graph nodes, $\mathcal{E}$ denotes relations among them, and $\mathcal{I}^{*}$ maps graph nodes to their supporting multimodal chunks. 
The evidence index connects the graph structure with the original document evidence.

\textbf{Online Graph-Guided Retrieval and Answering.}
Given a user query $q$, the online stage retrieves evidence through the constructed graph instead of directly searching over all chunks. 
The retriever first localizes query-relevant graph elements and then expands them to supporting chunks through $\mathcal{I}^{*}$:
\[
\mathcal{B}_q = \text{Retriever}(q, \mathcal{G}, \mathcal{I}^{*}).
\]
The retrieved chunks $\mathcal{B}_q$ are then used to generate the final answer:
\[
A = \text{LLM}(q, \mathcal{B}_q).
\]
Detailed task formulation is provided in Appendix~\ref{app:problem_statement}.

\begin{figure*}[!t]
  \includegraphics[width=1.0\linewidth]{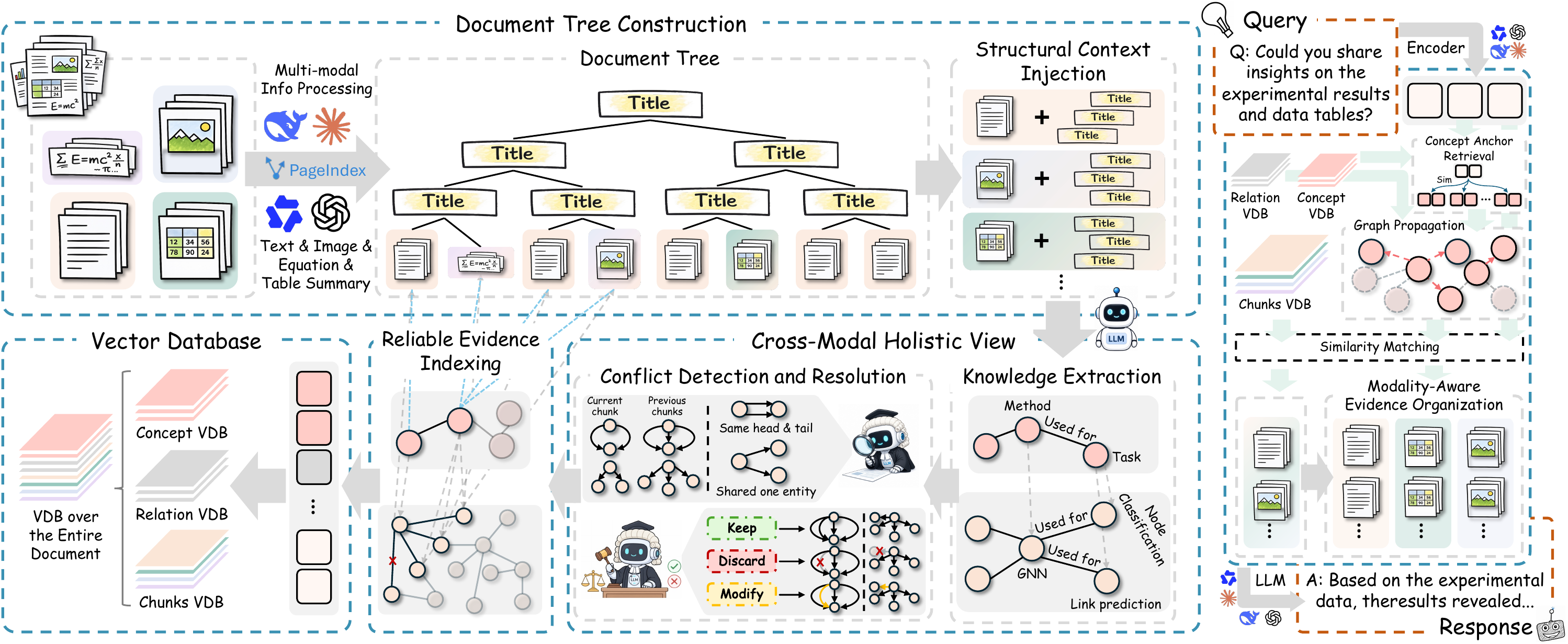}
  \vspace{-8mm}
  \caption{The offline graph construction and online retrieval of \ourmethod.
\textbf{(1)} A multimodal document tree provides structural context for each chunk.
\textbf{(2)} Holistic-view construction resolves cross-modal conflicts and builds a reliable concept-level index.
\textbf{(3)} Graph-guided retrieval searches the compact graph and organizes evidence by modality for answer generation.}
\Description{}\label{fig:framework}
\vspace{-4mm}
\end{figure*}

\subsection{Performance and Efficiency Analysis}
To better understand graph-based RAG over complex multimodal documents, we conduct a preliminary study on answer performance and retrieval efficiency. 
As shown in Fig.~\ref{fig:performance_and_query_time_between_different_model}, we compare representative graph-based RAG methods against the flat retrieval baseline BM25~\cite{robertson1994some} under the same evaluation setting. 
The comparison covers the text-based RAPTOR~\cite{sarthi2024raptor} and GraphRAG~\cite{edge2404local}. 
It also includes GraphRanker as a multimodal variant of HippoRAG~\cite{gutierrez2024hipporag}. 
The results show that graph structures do not always improve answer performance and can incur additional retrieval overhead.

\textbf{Performance Bottleneck.}
As shown in the left part of Fig.~\ref{fig:performance_and_query_time_between_different_model}, graph-based RAG methods do not consistently outperform the flat BM25 baseline. 
A possible cause is the absence of a Holistic View during graph construction. 
Knowledge extracted from local chunks may be inserted without being checked against existing graph elements or evidence from other modalities (Fig.~\ref{fig:flat_Multimodal_rag_and_Multimodal_graph_rag}(b)). 
This can produce unreliable graph-to-chunk indices that introduce incomplete, redundant, or conflicting evidence into the retrieved context.

\textbf{Efficiency Bottleneck.}
As shown in the right part of Fig.~\ref{fig:performance_and_query_time_between_different_model}, graph-based retrieval often requires more time than flat retrieval. 
This overhead arises from repeated node matching and relation traversal over dense graphs. 
Neighborhood expansion and evidence aggregation further increase the cost (Fig.~\ref{fig:flat_Multimodal_rag_and_Multimodal_graph_rag}(b)). 

These results show that effective Multimodal GraphRAG requires reliable evidence indexing and efficient retrieval over compact graph structures. 
This motivates our holistic-view-guided graph construction and graph-guided holistic retrieval.
\vspace{-1mm}
\section{Our Method}
To address the limitations of existing Multimodal GraphRAG methods~\cite{liu2025hm,luo2026hypergraphrag} in evidence
indexing reliability and retrieval efficiency, we propose \textbf{\ourmethod}, a
holistic-view multimodal GraphR-AG framework on complex document. 
The core idea is to build reliable evidence indices under holistic-view guidance and enable accurate and efficient retrieval over a compact concept graph.
As illustrated in Fig.\ref{fig:framework}, \ourmethod consists of two main modules: the Holistic-View-Guided Graph Construction Module (HGCM) and the
Graph-Guided Holistic Retrieval Module (GHRM). 

\subsection{Holistic-View-Guided Graph Construction}
Existing Multimodal GraphRAG methods often construct graphs from local chunks without holistic document-level and cross-modal context. 
This can produce incomplete, redundant, or conflicting indices, resulting in \textit{Unreliable Cross-Modal Evidence Indexing}~\cite{wu2026memgraphrag}.
To address this problem, \ourmethod constructs a concept-level indexing graph through three designs: (1) \textbf{Document Tree} provides chunks with document-level structural context; (2) \textbf{Cross-Modal Holistic View} $\mathcal{M}$ reduces noise, redundancy, and conflicts during graph construction; and (3) \textbf{Reliable Evidence Indexing via Entity-Level Graph Bridging} connects concept graph nodes directly to supporting multimodal chunks.
Together, these designs produce a compact and reliable knowledge indexing structure.

\subsubsection{Document Tree Construction}
A key reason for unreliable cross-modal evidence indexing is that chunks are often processed as isolated units during graph construction. 
Without the document-hierarchy context of a chunk, entities, values, and visual elements may be interpreted outside their intended section-level scope.
To alleviate this issue, \ourmethod organizes the input document into a multimodal document tree~\cite{wang2025bookrag,shin2026hikey,shin2025multidocfusion} that preserves both document hierarchy and modality information.
Given a document $D$, a layout parser first extracts a sequence of multimodal chunks:
\begin{align}
    \mathcal{B}
    =
    \operatorname{LayoutParser}(D)
    =
    \{b_i\}_{i=1}^{M},
    \quad
    b_i=(x_i,m_i),
\end{align}
where $\operatorname{LayoutParser}(\cdot)$ denotes the document layout parser, $M$ is the number of parsed chunks, and $x_i$ and $m_i$ denote the content and modality of $b_i$, respectively. 
These chunks are then organized into a document tree using the detected hierarchy and reading order:
\begin{align}
 \mathcal{T}=(\mathcal{N},\mathcal{E}_{\mathcal{T}})=\text{TreeBuilder}(\mathcal{B},D),   
\end{align}
where $\text{TreeBuilder}(\cdot)$ is an LLM-assisted tree constructor, $\mathcal{N}$ contains structural and chunk nodes, and $\mathcal{E}_{\mathcal{T}}$ contains parent--child edges. 
For each chunk $b_i$, $\text{Path}_{\mathcal{T}}(\text{root}, b_i)$ returns the ordered sequence of structural nodes from the document root to $b_i$, which is used as document-level structural context:
\begin{align}
    h_i=\text{Path}_{\mathcal{T}}(\text{root}, b_i).
\end{align}

This structural context specifies the document scope of each chunk, helping subsequent graph construction extract knowledge and build evidence indices under the correct section-level semantics.

\subsubsection{Cross-Modal Holistic View}
During indexing graph construction, knowledge extracted from multimodal chunks may be redundant or semantically conflicting with the existing graph~\cite{hong2024so,lee2025magic}. 
To improve the consistency of cross-modal knowledge extraction, \ourmethod maintains a Cross-Modal Holistic View $\mathcal{M}$ during graph construction. 
Here, $\mathcal{M}^{i-1}$ denotes the holistic view before processing chunk $b_i$, which contains the accepted facts from processed chunks.
Based on this view, \ourmethod performs conflict detection and resolution, thereby reducing noisy graph updates and improving the reliability of evidence indexing.

After constructing the document tree, \ourmethod uses the structural context of each chunk to support multimodal knowledge extraction. 
For each chunk $b_i=(x_i,m_i)$, its root-to-leaf path $h_i$ is concatenated with the chunk content $x_i$ and processed by a modality-specific extractor from a set of multimodal extractors:
\begin{align}
    (\mathcal{C}_i,\mathcal{U}_i,\mathcal{S}_i,\mathcal{F}_i)
    =
    \operatorname{Extractor}_{m_i}
    \left(
    h_i, x_i,
    \rho_{\mathrm{ext}}^{m_i}
    \right),
\end{align}
where $\mathcal{C}_i$, $\mathcal{U}_i$, $\mathcal{S}_i$, and $\mathcal{F}_i$ denote the sets of concepts, entities, schemas, and facts extracted from $b_i$, respectively. 
Both the extraction prompt $\rho_{\mathrm{ext}}^{m_i}$ and the extractor $\operatorname{Extractor}_{m_i}(\cdot)$ are selected based on the modality $m_i$ to accommodate different content.
The structural context $h_i$ helps the extractor interpret each chunk within its document-level scope, reducing extraction ambiguity.

The extracted facts are combined with those in the existing Cross-Modal Holistic View $\mathcal{M}^{i-1}$ to identify potential conflict groups:
\begin{align}
    \mathcal{F}^{+}_i = \mathcal{F}_i \cup \mathcal{M}^{i-1},
\end{align}
where $\mathcal{F}_i$ denotes the facts extracted from the current chunk $b_i$, $\mathcal{M}^{i-1}$ contains the accepted facts accumulated from all chunks processed prior to $b_i$, and $\mathcal{F}^{+}_i$ represents their combined fact set for conflict candidate construction.
For each fact $f=(e_h,r,e_t)$, we define three matching keys: $k_{hr}(f)=(e_h,r)$, $k_{ht}(f)=(e_h,e_t)$, and $k_h(f)=e_h$. 
The group of a newly extracted fact $f$ under key $k$ is:
\begin{align}
    g_k(f)
    =
    \{
    f'\in\mathcal{F}_i^{+}
    \mid
    k(f')=k(f)
    \}.
\end{align}

The set of potential conflict groups is defined as:
\begin{align}
    \mathcal{F}_{\mathrm{conf}}^{i}
    =
    \left\{
    g_k(f)
    \mid
    f\in\mathcal{F}_i,\ 
    k\in\{k_{hr},k_{ht},k_h\},\
    |g_k(f)|>1
    \right\}.
\end{align}


Each retained group contains at least one fact from the current chunk and captures a potential intra-chunk conflict or a conflict with previously accepted facts. 
However, structural overlap does not necessarily indicate a factual contradiction~\cite{jiayang2024econ,huang2025can}. 
Therefore, an LLM-based detector examines each candidate group and retains only confirmed conflicts:
\begin{align}
    \widehat{\mathcal{F}}_{\mathrm{conf}}^{i}
    =
    \operatorname{Detector}
(
\{g\mid g\in\mathcal{F}_{\mathrm{conf}}^i\},
\rho_{\mathrm{det}}
),
\end{align}
where $\rho_{\mathrm{det}}$ denotes the conflict-detection prompt and
$\widehat{\mathcal{F}}_{\mathrm{conf}}^{i}$ contains the confirmed conflict groups.

To support evidence-based resolution, we collect the entity-level knowledge nodes involved in each confirmed conflict group:
\begin{align}
    \mathcal{Z}_{g}
    =
    \{e_h,e_t \mid (e_h,r,e_t)\in g\}.
\end{align}

Their supporting chunks are then accessed through the entity-level evidence index:
\begin{align}
    \mathcal{B}_{g}
    =
    \bigcup_{e\in\mathcal{Z}_{g}}
    \mathcal{I}_{\mathrm{fac}}(e),
    \quad
    g\in\widehat{\mathcal{F}}_{\mathrm{conf}}^{i},
\end{align}
where $e$ denotes an entity-level node involved in the conflict group. 
The function $\mathcal{I}_{\mathrm{fac}}(\cdot)$ maps each entity-level knowledge node to its supporting chunks. 
For nodes newly extracted from $b_i$, the current chunk $b_i$ is recorded as their supporting chunk.

The confirmed conflicts and their supporting chunks are subsequently passed to an LLM-based resolver:
\begin{align}
    \mathcal{M}^i,\mathcal{I}_{\mathrm{fac}}^i
    =
    \operatorname{Resolver}
    (
    \mathcal{M}^{i-1},
    \mathcal{I}_{\mathrm{fac}}^{i-1},
    \mathcal{F}_i,
    \{(g,\mathcal{B}_g)\mid
    g\in\widehat{\mathcal{F}}_{\mathrm{conf}}^i\},
    b_i,
    \rho_{\mathrm{res}}
    ),
\end{align}
where $\mathcal{M}^i$ and $\mathcal{I}_{\mathrm{fac}}^i$ denote the updated holistic view and entity-level evidence index after processing $b_i$, respectively.
The resolver updates the evidence index accordingly: links supported only by discarded facts are removed, while those associated with retained, revised, or newly accepted facts are preserved or updated.

Based on the supporting evidence, the resolver merges redundant facts and revises inconsistent ones. 
It also retains complementary facts and removes unsupported ones. 
Facts outside the confirmed conflict groups in $\mathcal{F}_i$ are directly added to $\mathcal{M}^{i}$. 
The updated holistic view then guides the processing of subsequent chunks.

\subsubsection{Reliable Evidence Indexing via Entity-Level Graph Bridging}
To establish direct links from concept nodes to supporting chunks,
HVM-GraphRAG uses the resolved entity-level facts in $\mathcal{M}$
as an intermediate bridge.
Let $\mathcal{V}_{\mathrm{ent}}$ denote the set of entity nodes
induced by these facts.
Each entity $e\in\mathcal{V}_{\mathrm{ent}}$ is assigned to a unique
concept $\phi(e)$.
A concept-level graph
$\mathcal{G}_{\mathrm{con}}
=(\mathcal{V}_{\mathrm{con}},\mathcal{E}_{\mathrm{con}})$
is constructed as:
\begin{align}
    \mathcal{V}_{\mathrm{con}}
    &=
    \{\phi(e)\mid e\in\mathcal{V}_{\mathrm{ent}}\},\\
    \mathcal{E}_{\mathrm{con}}
    &=
    \{
    (\phi(e_h),r,\phi(e_t))
    \mid
    (e_h,r,e_t)\in\mathcal{M}
    \}.
\end{align}

For each concept node $c\in\mathcal{V}_{\mathrm{con}}$, its entity-level preimage is:
\begin{align}
    \phi^{-1}(c)
    =
    \{
    e\in\mathcal{V}_{\mathrm{ent}}
    \mid
    \phi(e)=c
    \}.
\end{align}

The concept-level evidence index is then constructed as:
\begin{align}
    \mathcal{I}_{\mathrm{con}}(c)
    =
    \bigcup_{e\in\phi^{-1}(c)}
    \mathcal{I}_{\mathrm{fac}}(e),
    \quad
    c\in\mathcal{V}_{\mathrm{con}},
\end{align}
where $\mathcal{I}_{\mathrm{fac}}(e)$ maps each entity node $e$
to its supporting multimodal chunks.
Therefore, $\mathcal{I}_{\mathrm{con}}(c)$ aggregates the evidence
associated with all entity nodes assigned to concept $c$, enabling
direct concept-to-evidence access.

This index enables direct evidence access over the compact concept graph without traversing the entity-level graph.

\subsection{Graph-Guided Holistic Retrieval}
During online retrieval, existing Multimodal GraphRAG methods often search dense entity-level graphs through node matching, relation traversal, and neighborhood expansion. 
As complex documents contain numerous entities, relations, and cross-modal evidence links, this process leads to \textit{Overly Expensive Graph Traversal}~\cite{hu2026remindrag,li2026scout}.
To reduce this cost, \ourmethod retrieves over the compact concept-level indexing graph
$\mathcal{G}_{\mathrm{con}}=(\mathcal{V}_{\mathrm{con}},\mathcal{E}_{\mathrm{con}})$ constructed offline.
The online process comprises three stages: (1) \textbf{Concept Anchor Retrieval} selects query-relevant concepts as retrieval anchors; (2) \textbf{Similarity-Based Graph Propagation} expands them through query-relevant relations; and (3) \textbf{Modality-Aware Evidence Organization} maps the selected elements to supporting chunks and organizes them by modality for answer generation.

\subsubsection{Concept Anchor Retrieval}
Rather than directly searching over all entity-level facts or multimodal chunks, \ourmethod first localizes the query within the compact concept-level graph. 
Concept nodes capture high-level semantics shared by heterogeneous document content, making them suitable retrieval anchors for queries whose supporting evidence may span multiple entities and modalities.
Given a query $q$, an LLM-based encoder $\operatorname{Enc}(\cdot)$ maps the query and each concept node $c \in \mathcal{V}_{con}$ into a shared semantic space~\cite{reimers2019sentence,karpukhin2020dense}, and their relevance is computed as:
\begin{align}
    s(q,c)
    =
    \operatorname{sim}
    (\operatorname{Enc}(q),\operatorname{Enc}(c)),
\end{align}
where $\operatorname{sim}(\cdot,\cdot)$ denotes cosine similarity. 
The top-$k_c$ concept nodes with high relevance scores are selected as the initial anchors:
\begin{align}
    \mathcal{A}_q
=
\{c \mid (c,s(q,c)) \in \operatorname{TopK}_{k_c}(\mathcal{V}_{con};s),\ s(q,c)\geq\theta\},
\end{align}
where $\mathcal{A}_q$ denotes the concept-anchor set, $\operatorname{TopK}_{k_c}(\mathcal{V}_{con};s)$ selects the top-$k_c$ concepts ranked by $s(q,c)$, and $\theta$ filters low-confidence concepts.
These anchors identify the most relevant semantic regions of the graph and constrain subsequent propagation to query-related structures, reducing unnecessary traversal.

\subsubsection{Similarity-Based Graph Propagation}

Although these anchors capture concepts directly related to the query, they may still miss related concepts needed for complete evidence retrieval. 
To expand the retrieval scope, \ourmethod measures the semantic similarity $s(q,r)$ between the query $q$ and each relation $r$ in the concept-level graph.
The top-$k_r$ relations with the highest similarity scores form the candidate relation set $\mathcal{R}_q$. 
To keep propagation constrained by reliable anchors, only candidate relations connected to at least one concept in $\mathcal{A}_q$ are retained:
\begin{align}
    \widehat{\mathcal{R}}_q
    =
    \{
    r \in \mathcal{R}_q
    \mid
    \exists (c_h,r,c_t)\in\mathcal{E}_{con},
    \{c_h,c_t\}\cap\mathcal{A}_q\neq\emptyset
    \},
\end{align}
where $\mathcal{E}_{con}$ denotes the set of concept-level edges in $\mathcal{G}_{con}$, and each edge is a triplet $(c_h,r,c_t)$. 
The concepts reached through the retained relations are then merged with the initial anchors:
\begin{align}
    \widehat{\mathcal{A}}_q
    =
    \mathcal{A}_q
    \cup
    \{c_h,c_t \mid
    (c_h,r,c_t)\in\mathcal{E}_{con},
    r\in\widehat{\mathcal{R}}_q,
    \{c_h,c_t\}\cap\mathcal{A}_q\neq\emptyset
    \}.
\end{align}

This constrained propagation expands the retrieved concept set through query-relevant relations while avoiding propagation toward unrelated graph regions.

\subsubsection{Modality-Aware Evidence Organization}
Retrieved evidence from complex documents often spans text, tables, and images. 
Interleaving these heterogeneous chunks may obscure modality-specific semantics and hinder effective evidence integration~\cite{wang2026meg}. 
To address this issue, \ourmethod first constructs a candidate evidence set relevant to the query and then organizes the selected chunks into modality-specific groups before answer generation.
{
\setlength{\dashlinedash}{3pt}
\setlength{\dashlinegap}{2pt}
\setlength{\arrayrulewidth}{0.5pt}

\begin{table}[b]
\centering
\vspace{-4mm}
\caption{Statistics of the datasets.}
\vspace{-4mm}
\label{tab:dataset_statistics}
\scriptsize
\setlength{\tabcolsep}{3.0pt}
\renewcommand{\arraystretch}{1.}
\resizebox{0.99\columnwidth}{!}{
\begin{tabular}{cccc}
\toprule
\textbf{Dataset} & \textbf{MMLongBench} & \textbf{M3DocVQA} & \textbf{Qasper} \\
\midrule
Questions   & 669       & 633       & 640 \\
\noalign{\vskip 1pt}
\cdashline{1-4}
\noalign{\vskip 1pt}
Documents   & 85        & 500       & 192 \\
\noalign{\vskip 1pt}
\cdashline{1-4}
\noalign{\vskip 1pt}
Avg. Pages  & 42.16     & 8.52      & 10.95 \\
\noalign{\vskip 1pt}
\cdashline{1-4}
\noalign{\vskip 1pt}
Avg. Images & 25.92     & 3.51      & 3.43 \\
\noalign{\vskip 1pt}
\cdashline{1-4}
\noalign{\vskip 1pt}
\makecell[c]{Cross-Modal\\Questions (\%)} & 22.73 & 0.00 & 8.31 \\
\noalign{\vskip 1pt}
\cdashline{1-4}
\noalign{\vskip 1pt}
Tokens      & 2,816,155 & 3,553,774 & 2,265,349 \\
\noalign{\vskip 1pt}
\cdashline{1-4}
\noalign{\vskip 1pt}
Metrics     & EM, F1    & EM, F1    & Acc, F1 \\
\bottomrule
\end{tabular}
}
\end{table}
}

\begin{table*}[!t]
\centering
\caption{Performance comparison on three complex document QA datasets (MMLongBench, M3DocVQA and Qasper). Best and second-best results are shown in bold and underlined, respectively.}
\label{tab:overall_performance}
\vspace{-4mm}
\small
\resizebox{\textwidth}{!}{
\begin{tabular}{clcccccc}
\toprule
\multirow{2}{*}{\textbf{Baseline Type}} 
& \multirow{2}{*}{\textbf{Method}} 
& \multicolumn{2}{c}{\textbf{MMLongBench}} 
& \multicolumn{2}{c}{\textbf{M3DocVQA}} 
& \multicolumn{2}{c}{\textbf{Qasper}} \\
\cmidrule(lr){3-4} \cmidrule(lr){5-6} \cmidrule(lr){7-8}
& 
& \textbf{Exact Match} & \textbf{F1-score} 
& \textbf{Exact Match} & \textbf{F1-score} 
& \textbf{Accuracy} & \textbf{F1-score} \\
\midrule

\multirow{3}{*}{\makecell[c]{Conventional\\RAG}}
& \cellcolor{darkgray!10}BM25             & \cellcolor{lightgray!25}18.3 & \cellcolor{lightgray!25}20.2 & 3\cellcolor{lightgray!25}4.6 & \cellcolor{lightgray!25}37.8 & \cellcolor{lightgray!25}38.1 & \cellcolor{lightgray!25}42.5 \\
& \cellcolor{darkgray!10}Vanilla RAG      & \cellcolor{lightgray!25}16.5$\textcolor{subCyan}{_{\downarrow 1.8}}$ & \cellcolor{lightgray!25}18.0$\textcolor{subCyan}{_{\downarrow 2.2}}$ & \cellcolor{lightgray!25}36.5$\textcolor{subRed}{_{\uparrow 1.9}}$ & \cellcolor{lightgray!25}40.2$\textcolor{subRed}{_{\uparrow 2.4}}$ & \cellcolor{lightgray!25}40.6$\textcolor{subRed}{_{\uparrow 2.5}}$ & \cellcolor{lightgray!25}44.4$\textcolor{subRed}{_{\uparrow 1.9}}$ \\
& \cellcolor{darkgray!10}Layout + Vanilla & \cellcolor{lightgray!25}18.1$\textcolor{subCyan}{_{\downarrow 0.2}}$ & \cellcolor{lightgray!25}19.8$\textcolor{subCyan}{_{\downarrow 0.4}}$ & \cellcolor{lightgray!25}36.9$\textcolor{subRed}{_{\uparrow 2.3}}$ & \cellcolor{lightgray!25}40.2$\textcolor{subRed}{_{\uparrow 2.4}}$ & \cellcolor{lightgray!25}40.7$\textcolor{subRed}{_{\uparrow 2.6}}$ & \cellcolor{lightgray!25}44.6$\textcolor{subRed}{_{\uparrow 2.1}}$ \\
\noalign{\vskip 2pt}
\cdashline{1-8}
\noalign{\vskip 2pt}
\multirow{3}{*}{GraphRAG}
& \cellcolor{darkpurple!25}RAPTOR           & \cellcolor{lightpurple!25}21.3$\textcolor{subRed}{_{\uparrow 3.0}}$ & \cellcolor{lightpurple!25}21.8$\textcolor{subRed}{_{\uparrow 1.6}}$ & \cellcolor{lightpurple!25}34.3$\textcolor{subCyan}{_{\downarrow 0.3}}$ & \cellcolor{lightpurple!25}37.3$\textcolor{subCyan}{_{\downarrow 0.5}}$ & \cellcolor{lightpurple!25}39.4$\textcolor{subRed}{_{\uparrow 1.3}}$ & \cellcolor{lightpurple!25}44.1$\textcolor{subRed}{_{\uparrow 1.6}}$ \\
& \cellcolor{darkpurple!25}GraphRAG-Local   & \cellcolor{lightpurple!25}7.7$\textcolor{subCyan}{_{\downarrow 10.6}}$  & \cellcolor{lightpurple!25}8.5$\textcolor{subCyan}{_{\downarrow 11.7}}$  & \cellcolor{lightpurple!25}23.7$\textcolor{subCyan}{_{\downarrow 10.9}}$ & \cellcolor{lightpurple!25}25.6$\textcolor{subCyan}{_{\downarrow 12.2}}$ & \cellcolor{lightpurple!25}35.9$\textcolor{subCyan}{_{\downarrow 2.2}}$ & \cellcolor{lightpurple!25}39.2$\textcolor{subCyan}{_{\downarrow 3.3}}$ \\
& \cellcolor{darkpurple!25}GraphRAG-Global  & \cellcolor{lightpurple!25}5.3$\textcolor{subCyan}{_{\downarrow 13.0}}$  & \cellcolor{lightpurple!25}5.6$\textcolor{subCyan}{_{\downarrow 14.6}}$  & \cellcolor{lightpurple!25}20.2$\textcolor{subCyan}{_{\downarrow 14.4}}$ & \cellcolor{lightpurple!25}22.0$\textcolor{subCyan}{_{\downarrow 15.8}}$ & \cellcolor{lightpurple!25}24.0$\textcolor{subCyan}{_{\downarrow 14.1}}$ & \cellcolor{lightpurple!25}24.1$\textcolor{subCyan}{_{\downarrow 18.4}}$\\
\noalign{\vskip 2pt}
\cdashline{1-8}
\noalign{\vskip 2pt}
\multirow{3}{*}{\makecell[c]{Layout segmented\\RAG}}
& \cellcolor{darkgreen!25}MM-Vanilla       & \cellcolor{lightgreen!25}6.8$\textcolor{subCyan}{_{\downarrow 11.5}}$  & \cellcolor{lightgreen!25}8.4$\textcolor{subCyan}{_{\downarrow 11.8}}$  & \cellcolor{lightgreen!25}25.1$\textcolor{subCyan}{_{\downarrow 9.5}}$ & \cellcolor{lightgreen!25}27.7$\textcolor{subCyan}{_{\downarrow 10.1}}$ & \cellcolor{lightgreen!25}27.9$\textcolor{subCyan}{_{\downarrow 10.2}}$ & \cellcolor{lightgreen!25}29.3$\textcolor{subCyan}{_{\downarrow 13.2}}$ \\
& \cellcolor{darkgreen!25}Tree-Traverse    & \cellcolor{lightgreen!25}12.7$\textcolor{subCyan}{_{\downarrow 5.6}}$ & \cellcolor{lightgreen!25}14.4$\textcolor{subCyan}{_{\downarrow 5.8}}$ & \cellcolor{lightgreen!25}33.3$\textcolor{subCyan}{_{\downarrow 1.3}}$ & \cellcolor{lightgreen!25}36.2$\textcolor{subCyan}{_{\downarrow 1.6}}$ & \cellcolor{lightgreen!25}27.3$\textcolor{subCyan}{_{\downarrow 10.8}}$ & \cellcolor{lightgreen!25}32.1$\textcolor{subCyan}{_{\downarrow 10.4}}$ \\
& \cellcolor{darkgreen!25}DocETL           & \cellcolor{lightgreen!25}27.5$\textcolor{subRed}{_{\uparrow 9.2}}$ & \cellcolor{lightgreen!25}28.6$\textcolor{subRed}{_{\uparrow 8.4}}$ & \cellcolor{lightgreen!25}40.9$\textcolor{subRed}{_{\uparrow 6.3}}$ & \cellcolor{lightgreen!25}43.3$\textcolor{subRed}{_{\uparrow 5.5}}$ & \cellcolor{lightgreen!25}42.3$\textcolor{subRed}{_{\uparrow 4.2}}$ & \cellcolor{lightgreen!25}50.4$\textcolor{subRed}{_{\uparrow 7.9}}$ \\
\noalign{\vskip 2pt}
\cdashline{1-8}
\noalign{\vskip 2pt}
\multirow{2}{*}{\makecell[c]{Multimodal\\GraphRAG}}
& \cellcolor{darkyellow!25}GraphRanker      & \cellcolor{lightyellow!25}21.2$\textcolor{subRed}{_{\uparrow 2.9}}$ & \cellcolor{lightyellow!25}22.7$\textcolor{subRed}{_{\uparrow 2.5}}$ & \cellcolor{lightyellow!25}43.0$\textcolor{subRed}{_{\uparrow 8.4}}$ & \cellcolor{lightyellow!25}47.8$\textcolor{subRed}{_{\uparrow 10.0}}$ & \cellcolor{lightyellow!25}32.9$\textcolor{subCyan}{_{\downarrow 5.2}}$ & \cellcolor{lightyellow!25}37.6$\textcolor{subCyan}{_{\downarrow 4.9}}$ \\
& \cellcolor{darkyellow!25}BookRAG
& \cellcolor{lightyellow!25}\underline{43.8}$\textcolor{subRed}{_{\uparrow 25.5}}$ & \cellcolor{lightyellow!25}\underline{44.9}$\textcolor{subRed}{_{\uparrow 24.7}}$
& \cellcolor{lightyellow!25}\underline{61.0}$\textcolor{subRed}{_{\uparrow 26.4}}$ & \cellcolor{lightyellow!25}\textbf{66.2}$\textcolor{subRed}{_{\uparrow 28.4}}$
& \cellcolor{lightyellow!25}\underline{55.2}$\textcolor{subRed}{_{\uparrow 17.1}}$ & \cellcolor{lightyellow!25}\underline{61.1}$\textcolor{subRed}{_{\uparrow 18.6}}$ \\
\noalign{\vskip 2pt}
\cdashline{1-8}
\noalign{\vskip 2pt}
\multirow{1}{*}{\textbf{Our proposed}} & \cellcolor{darkred!25}\textbf{\ourmethod} & \cellcolor{lightred!25}\textbf{54.1}$\textcolor{subRed}{_{\uparrow 35.8}}$ & \cellcolor{lightred!25}\textbf{62.8}$\textcolor{subRed}{_{\uparrow 42.6}}$ & \cellcolor{lightred!25}\textbf{63.0}$\textcolor{subRed}{_{\uparrow 28.4}}$ & \cellcolor{lightred!25}\underline{66.0}$\textcolor{subRed}{_{\uparrow 28.2}}$& \cellcolor{lightred!25}\textbf{66.1}$\textcolor{subRed}{_{\uparrow 28.0}}$& \cellcolor{lightred!25}\textbf{71.5}$\textcolor{subRed}{_{\uparrow 29.0}}$ \\
\bottomrule
\end{tabular}
}
\vspace{-4mm}
\end{table*}

After graph propagation, $\widehat{\mathcal{A}}_q$ and $\widehat{\mathcal{R}}_q$ define the query-relevant concept subgraph. 
\ourmethod directly accesses its supporting chunks through $\mathcal{I}_{\mathrm{con}}(\cdot)$, avoiding entity-level graph traversal:
\begin{align}
    \mathcal{B}_q^{\mathrm{graph}}
    =
    \bigcup_{c\in\widehat{\mathcal{A}}_q}
    \mathcal{I}_{\mathrm{con}}(c),
\end{align}
where $\mathcal{B}_q^{\mathrm{graph}}$ denotes the graph-guided candidate chunk set, $c$ denotes a concept-level graph node from $\widehat{\mathcal{A}}_q$, and $\mathcal{I}_{\mathrm{con}}(\cdot)$ maps each graph node to its supporting chunks.
Each graph-retrieved chunk $b\in\mathcal{B}_q^{\mathrm{graph}}$ is converted into a modality-specific textual representation $\operatorname{Rep}_{m_b}(b)$. 
This function preserves the original representations of text and table chunks. 
For image chunks, it uses a VLM to generate textual summaries.
The semantic similarity between the query $q$ and chunk $b$ is then computed as:
\begin{align}
    s(q,b)
    =
    \operatorname{sim}
    (
    \operatorname{Enc}(q),
    \operatorname{Enc}(\operatorname{Rep}_{m_b}(b))
    ).
\end{align}


The top-$K$ chunks are retained as the graph-guided evidence set $\mathcal{B}_q^{K}$.
Graph-guided retrieval localizes semantically connected evidence. 
Direct query--chunk matching supplements the evidence set with relevant chunks that may be overlooked during semantic graph construction.
The supplementary evidence set is obtained by retrieving the top-$k_b$ chunks from $\mathcal{B}$:
\begin{align}
    \mathcal{B}_q^{\mathrm{rag}}
    =
    \operatorname{TopK}_{k_b}
    \left(
    \{(b,s(q,b))\mid b\in\mathcal{B}\}
    \right).
\end{align}

The graph-guided evidence and conventional RAG evidence are then merged:
\begin{align}
    \mathcal{B}_q^{\mathrm{final}}
    =
    \mathcal{B}_q^{K}
    \cup
    \mathcal{B}_q^{\mathrm{rag}},
\end{align}
where $\mathcal{B}_q^{\mathrm{final}}$ denotes the final candidate chunk set before modality-aware organization.
Although $\mathcal{B}_q^{\mathrm{final}}$ contains relevant evidence, modality-interleaved chunks may make heterogeneous evidence harder for the answering model to compare and integrate.
To present the retrieved information more clearly, \ourmethod partitions $\mathcal{B}_q^{\mathrm{final}}$ by modality:
\begin{align}
    \mathcal{B}_q^m
    =
    \{b\in\mathcal{B}_q^{\mathrm{final}} \mid m_b=m\},
    \quad m\in\Omega,
\end{align}
where $\Omega=\{\text{text},\text{table},\text{image}\}$ denotes the modality set. 
Chunks in each modality group are ranked by $s(q,b)$. 
The top-$k_m$ chunks are retained to construct the final evidence context:
\begin{align}
    \mathcal{B}_q^{\mathrm{maeo}}
    =
    \bigcup_{m\in\Omega}
    \operatorname{TopK}_{k_m}(\mathcal{B}_q^m),
\end{align}
where $\mathcal{B}_q^{\mathrm{maeo}}$ denotes the modality-aware evidence set.

The selected chunks are arranged in a predefined modality order and provided to the answering model:
\begin{align}
    A=\operatorname{Answer}(q,\mathcal{B}_q^{\mathrm{maeo}},\rho_{\mathrm{ans}}),
\end{align}
where $\operatorname{Answer}(\cdot)$ and $\rho_{\mathrm{ans}}$ denote the answer-generation pipeline and prompt, respectively. 
Within this pipeline, a VLM converts each image chunk into a textual summary. 
The summaries and textual representations of non-image chunks are ordered by modality and passed to the LLM for answer generation.
This modality-specific organization enables the answering model to distinguish and integrate heterogeneous evidence more effectively.

\section{Experiments}
In this section, we conduct extensive experiments to answer the following research questions:

\begin{itemize}
    \item \textbf{RQ1:} How does \ourmethod perform against state-of-the-art methods on complex document QA?
    
    \item \textbf{RQ2:} How does \ourmethod compare with existing RAG methods in online latency and token consumption?
    
    \item \textbf{RQ3:} What extent does each component contribute to the overall performance of \ourmethod?
    
    \item \textbf{RQ4:} How do different hyperparameter settings affect the performance of \ourmethod?
\end{itemize}

Additional experimental results, including case studies and further analyses, are provided in the Appendix~\ref{app:exp_det}.

\subsection{Experimental Setting}
\subsubsection{Datasets}
We evaluate \ourmethod on three complex document QA datasets: MMLongBench~\cite{ma2024mmlongbench}, M3DocVQA~\cite{cho2024m3docrag}, and Qasper~\cite{dasigi2021dataset}. 
\textbf{MMLongBench} contains long documents from diverse domains, including guidebooks, financial reports, and industry documents. 
\textbf{M3DocVQA} comprises multimodal HTML-style documents derived from Wikipedia and requires QA across heterogeneous document elements. 
\textbf{Qasper} focuses on scientific papers, with supporting evidence often distributed throughout the document. 
We use only the processed benchmark versions of these datasets provided by prior work~\cite{wang2025bookrag}, without introducing any additional data filtering, question augmentation, or split modification.
The statistics of the three datasets are summarized in Table~\ref{tab:dataset_statistics}, with detailed descriptions provided in the Appendix~\ref{app:exp_dataset}.

\subsubsection{Baselines}
We compare \ourmethod with representative methods from four RAG categories. \textbf{Conventional RAG} includes BM25~\cite{robertson1994some}, Vanilla RAG, and Layout+Vanilla. \textbf{Graph-Based RAG} includes RAPTOR~\cite{sarthi2024raptor} and two variants of GraphRAG~\cite{edge2404local}: GraphRAG-Local and GraphRAG-Global. \textbf{Layout segmented RAG} includes MM-Vanilla, Tree-Traverse~\cite{zhang2025pageindex}, and DocETL~\cite{shankar2024docetl}. \textbf{Multimodal GraphRAG} includes GraphRanker~\cite{gutierrez2024hipporag} and BookRAG~\cite{wang2025bookrag}. Details of these methods are provided in the Appendix~\ref{exp:baselines}.

\subsubsection{Metrics}
We evaluate all methods using a unified answer-extraction and normalization pipeline and report Exact Match (EM), token-level F1 score, and string-based accuracy (Acc). EM measures the proportion of predictions that exactly match the reference answers. F1 evaluates token-level overlap between predicted and reference answers. Acc measures the proportion of correctly answered questions. Higher values indicate better performance for all three metrics.
Detailed definitions and calculation procedures for these metrics are provided in the Appendix~\ref{app:Eva_Metric}.

\subsubsection{Implementation Details}
For a fair comparison, all methods use the same embedding model and answer-generation model. 
We employ \texttt{Qwen3-Embedding-0.6B}~\cite{zhang2025qwen3} to encode queries and retrieval units. 
The document layout is parsed using \texttt{Mineru}~\cite{wang2024mineru}. 
During graph construction, \texttt{Qwen3-8B-AWQ}~\cite{yang2025qwen3} extracts knowledge from textual chunks while \texttt{Qwen2.5-VL}~\cite{wu2025qwen} processes visual chunks. 
The same \texttt{Qwen3-8B-AWQ} is used for conflict resolution and final answer generation.
For online retrieval, we select the $K=10$ for top-$K$ retrieval in all methods.
And we set the temperature to 0 for all LLM and VLM calls to ensure reproducibility. 
The experiments are conducted on an Intel(R) Xeon(R) Gold 5120 CPU and four NVIDIA L40 GPUs with 48 GB of memory each.
Further implementation details and hyperparameter settings are provided in the Appendix~\ref{app:imp_details}.

\subsection{Overall Performance (RQ1)}
To answer RQ1, we compare the answer generation performance of \ourmethod with all baseline methods on MMLongBench, M3DocVQA, and Qasper. The results are reported in Table~\ref{tab:overall_performance}. Based on these results, we make the following observations:

\textbf{Graph-based RAG shows inconsistent performance gains.}
RAPTOR achieves 39.4\% Accuracy on Qasper and outperforms BM25 by 1.3 percentage points.
In contrast, GraphRAG-Local and GraphRAG-Global achieve 35.9\% and 24.0\% Accuracy. 
They underperform BM25 by 2.2 percentage points and 14.1 percentage points, respectively. 
These results suggest that graph structures alone do not guarantee better retrieval.
A possible explanation is that unreliable graph-to-evidence associations introduce conflicting information into the retrieved context.

\textbf{Layout segmented RAG is sensitive to model design.}
DocETL consistently performs best within this category.
It achieves Exact Match of 27.5\% on MMLongBench dataset.
In comparison, Tree-Traverse obtains 12.7\% while MM-Vanilla reaches only 6.8\%.
This gap may arise from their different uses of document structure.
Unlike simple multimodal retrieval or tree navigation, DocETL performs structured document transformations before retrieval.
This suggests that layout information requires an effective processing mechanism to improve QA performance.

\textbf{Multimodal GraphRAG achieves inconsistent gains.}
Graph-Ranker achieves 37.6\% F1 on Qasper and trails BM25 by 4.9\%.
In contrast, BookRAG achieves F1-scores of 61.1\% on the Qasper dataset. 
This gap shows that multimodal graph construction alone does not guarantee strong QA performance. 
BookRAG benefits from hierarchical document organization and adaptive retrieval. 
However, neither method explicitly verifies the reliability of graph-to-evidence indices. 
Noisy or conflicting evidence associations may therefore limit their retrieval quality.

\textbf{\ourmethod achieves the best result on five of the six reported metrics.}
These gains are consistent with the intended benefit of holistic-view-guided graph construction.
The improvements are particularly large on MMLongBench and Qasper with F1-score gains of 17.9 and 10.4 percentage points over BookRAG.
Cross-modal questions account for 22.73\% and 8.31\% of these datasets.
By contrast, M3DocVQA contains no such questions and shows only limited improvement.
This pattern suggests that the gains of \ourmethod may be more pronounced when answering requires cross-modal evidence integration.

\subsection{Efficiency Comparison (RQ2)}
To answer RQ2, we compare the online query efficiency of \ourmethod with baseline methods by reporting total query time and token consumption across all evaluation questions on MMLongBench and Qasper.
All measurements are collected after document preprocessing and indexing and therefore exclude offline construction costs.
The results are presented in Fig.~\ref{fig:efficiency_and_cost}, with complete results and further analyses provided in Appendix~\ref{app:eff_comp}.
Based on these results, we make the following observations:

\textbf{Graph-based and layout-segmented RAG methods generally incur higher retrieval costs.}
Graph-based methods require matching and traversal over complex graph structures.
Layout-segmented methods also process more structural units with richer content.
Consequently, most of these methods consume more query time and tokens than conventional RAG on both datasets.
BookRAG reduces some graph-related overhead but still relies on hierarchical traversal and LLM-driven adaptive retrieval.
Its efficiency therefore remains less competitive than conventional RAG.

\begin{figure}[!t]
  \includegraphics[width=1.0\linewidth]{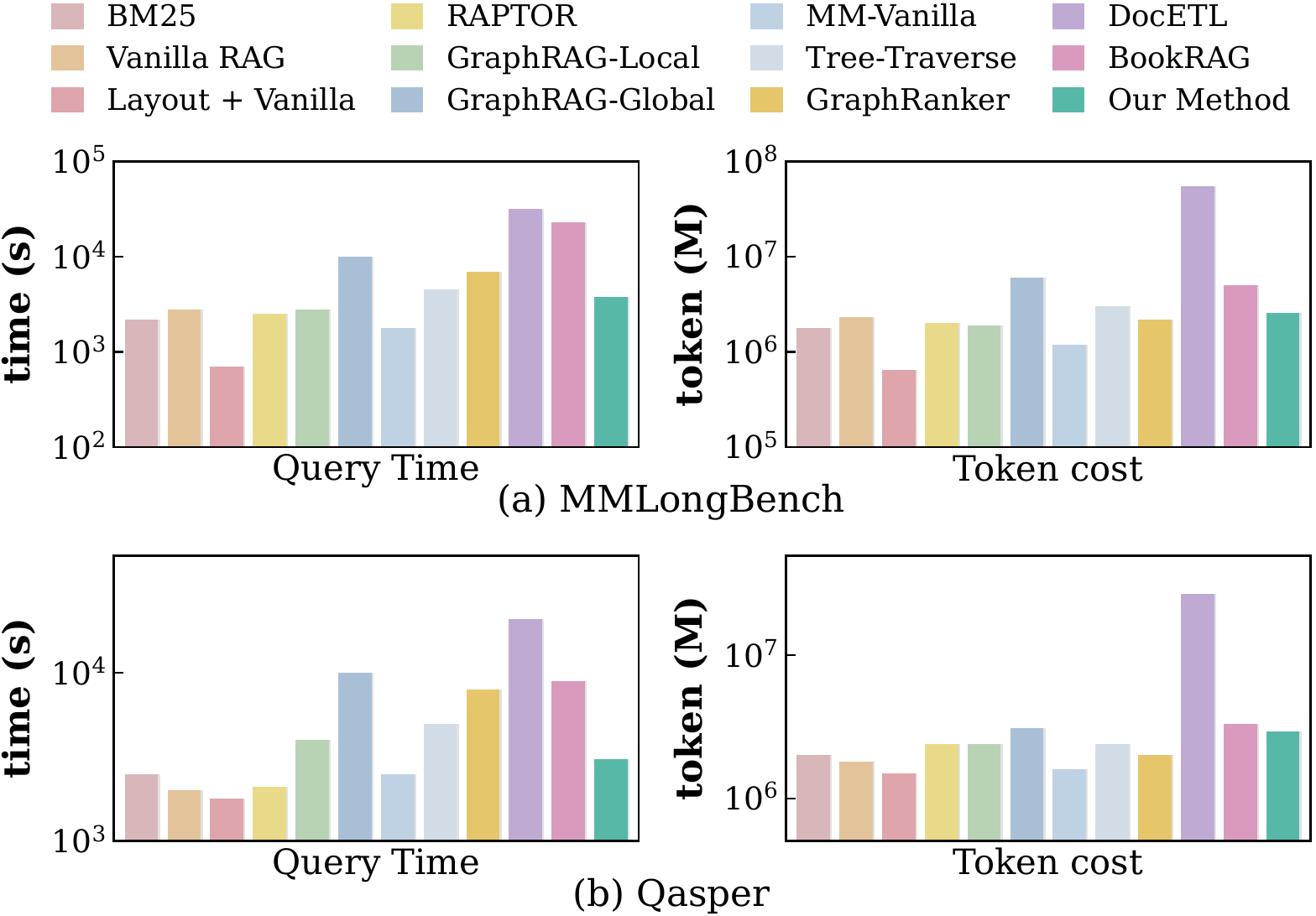}
  \vspace{-6mm}
  \caption{Comparison of query efficiency.}\label{fig:efficiency_and_cost}
\vspace{-4mm}
\end{figure}

\begin{figure}[!t]
  \includegraphics[width=1.0\linewidth]{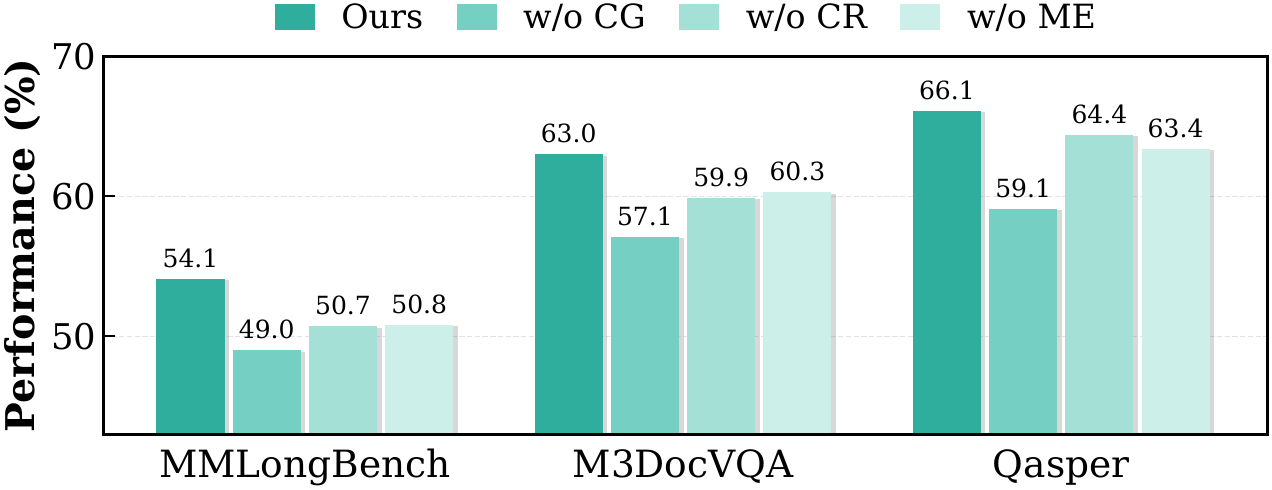}
  \vspace{-6mm}
  \caption{Ablation results on MMLongBench/M3DocVQA (EM) and Qasper (Acc).}\label{fig:ablation_study}
\vspace{-6mm}
\end{figure}

\textbf{\ourmethod achieves efficiency comparable to conventional RAG.}
Our proposed \ourmethod maintains low total query time and token consumption
on both datasets.
This efficiency comes from retrieving over a compact concept-level graph and directly mapping selected graph nodes to supporting chunks.
Moreover, the retrieval process requires no additional LLM-based query analysis.
These designs avoid entity-level graph traversal and substantially reduce online retrieval costs.

\subsection{Ablation Study (RQ3)}
To answer RQ3, we conduct ablation studies on all three datasets to evaluate the contribution of each component. We compare the full model with three variants that remove the \textit{Concept-Guided Graph} (w/o CG), \textit{Conflict Resolution} (w/o CR), and \textit{Modality-Aware Evidence Organization} (w/o ME), respectively. As shown in Fig.\ref{fig:ablation_study}, the full model consistently achieves the best performance of 54.1\%, 63.0\%, and 66.1\%. Removing any component leads to performance degradation across all datasets. 
These results demonstrate that the three components jointly improve evidence retrieval and final QA performance.

\textbf{w/o Concept-Guided Graph.}
Removing the Concept-Guided Graph leads to the largest performance decline with scores dropping to 49.0\%, 57.1\%, and 59.1\%. Without compact concept-level retrieval anchors, the model cannot effectively localize relevant graph regions and may retrieve less relevant evidence.

\textbf{w/o Conflict Resolution.}
Removing Conflict Resolution causes consistent degradation across all datasets and yields scores of 50.7\%, 59.9\%, and 64.4\%. Without global conflict detection and resolution, redundant or contradictory facts remain in the graph and weaken the reliability of evidence indexing.

\textbf{w/o Modality-Aware Evidence Organization.}
Excluding Mod-ality-Aware Evidence Organization produces scores of 50.8\%, 60.3\%, and 63.4\%. Without modality-specific organization, heterogeneous chunks remain interleaved in the retrieved context and increase the difficulty for the answering model to interpret and integrate multimodal evidence.

\begin{figure}[t]
\centering
{
{\includegraphics[width=0.49\linewidth]{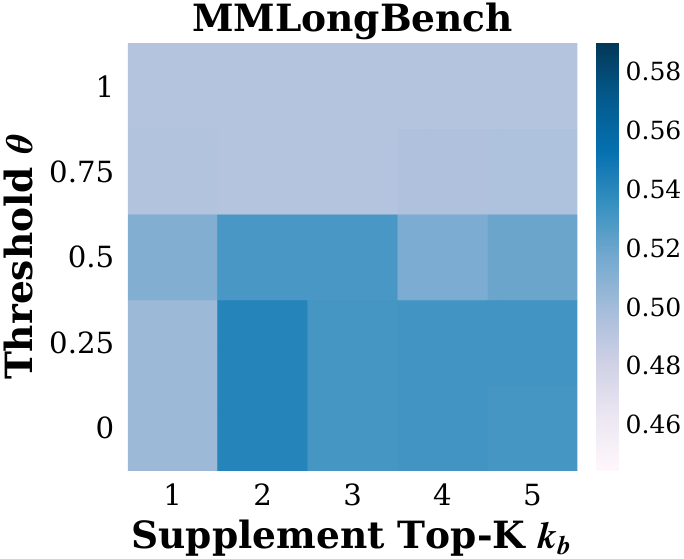}}}
{
{\includegraphics[width=0.49\linewidth]{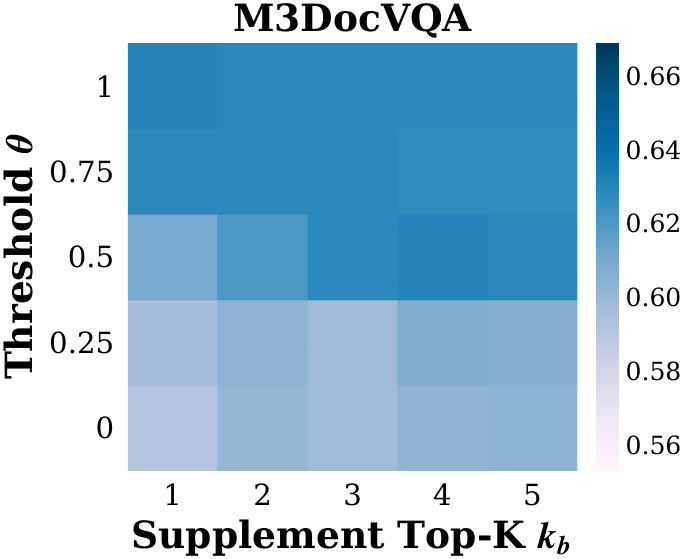}}}%
\vspace{-2mm}
\Description{The edge degree distribution}
\caption{Effects of the similarity threshold $\theta$ and the number of supplementary chunks $k_b$ on MMLongBench and M3DocVQA. Darker colors indicate higher performance.}\label{fig:threshold_and_supp}
\vspace{-2mm}
\end{figure}

\begin{figure}[t]
\centering
{
{\includegraphics[width=0.49\linewidth]{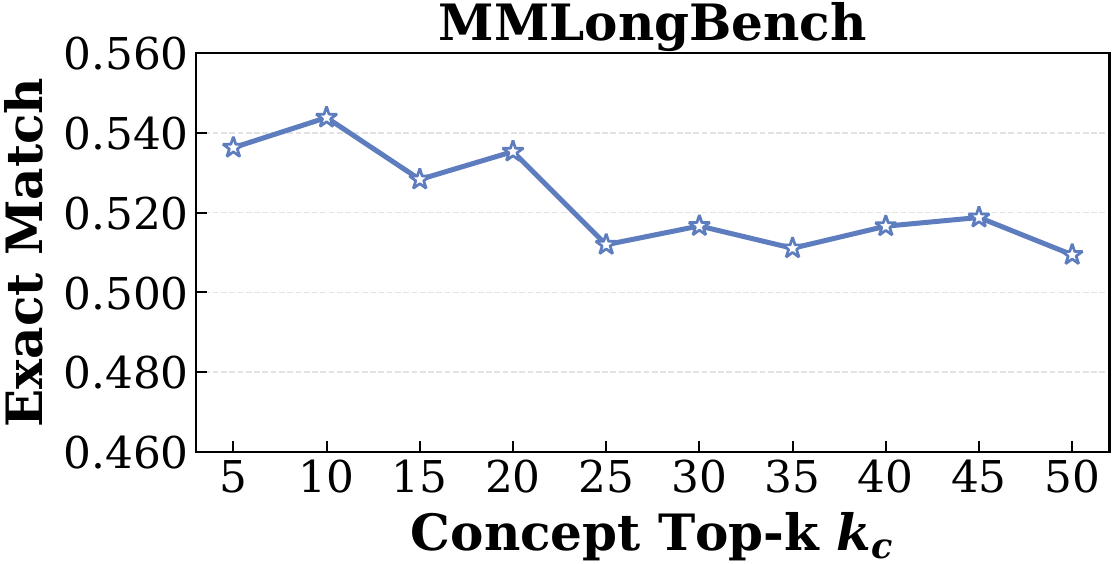}}}%
{
{\includegraphics[width=0.49\linewidth]{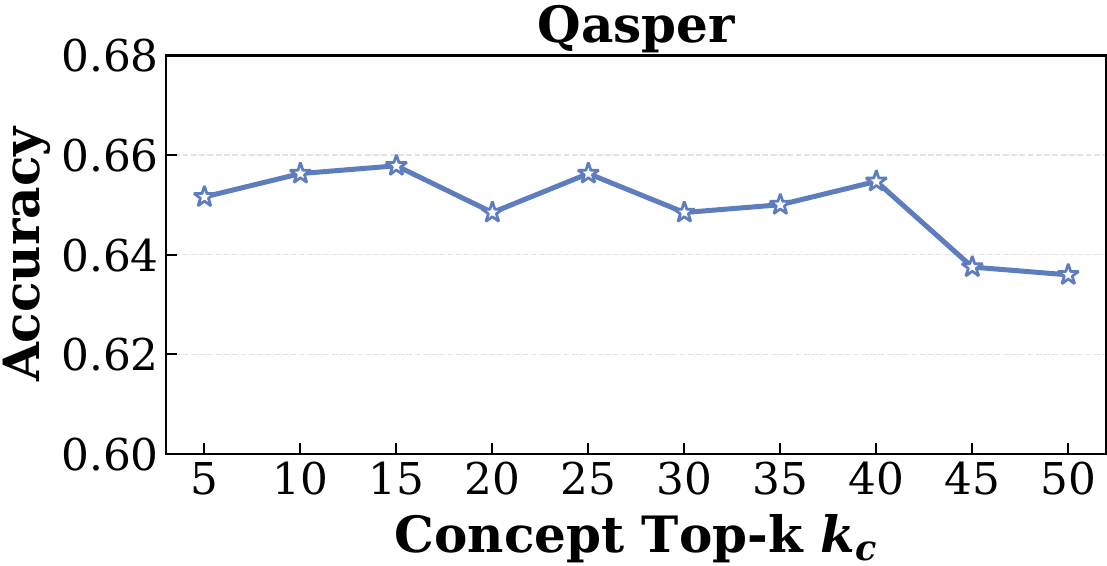}}}%
\vspace{-2mm}
\Description{The edge degree distribution}
\caption{Effects of the number of retrieved concepts $k_c$ on MMLongBench and Qasper.}\label{fig:concept_top_k}
\vspace{-4mm}
\end{figure}

\subsection{Hyperparameter Analysis (RQ4)}
To answer RQ4, we examine two groups of retrieval hyperparameters.
First, we jointly analyze the similarity threshold $\theta$ for filtering concept anchors and the number $k_b$ of supplementary chunks retrieved by conventional RAG method.
Second, we vary the number $k_c$ of top-ranked concept nodes selected from the compact concept graph. All remaining parameters are fixed during each experiment. The results are presented in Fig.\ref{fig:threshold_and_supp} and Fig.\ref{fig:concept_top_k}.

\textbf{Effects of $\theta$ and $k_b$.}
As shown in Fig.~\ref{fig:threshold_and_supp}, the influence of $\theta$ varies across datasets. 
On MMLongBench, lower thresholds generally achieve stronger performance, especially around $\theta=0.25$ and $k_b=2$. 
This suggests that retaining more concept anchors helps cover distributed evidence in long documents. 
In contrast, M3DocVQA performs better under higher thresholds, such as $\theta=0.5$. 
Since M3DocVQA contains more localized visual question answering cases, stricter concept-anchor filtering can reduce unnecessary graph expansion and keep retrieval focused on highly matched evidence. 
The effect of $k_b$ is also dataset-dependent. 
MMLongBench benefits from a small number of supplementary chunks, while M3DocVQA remains relatively stable when $k_b$ increases. 
Overall, these results show that properly setting $\theta$ and $k_b$ helps \ourmethod balance graph-guided retrieval and local evidence supplementation, thereby improving the final QA performance.

\textbf{Effects of $k_c$.}
As shown in Fig.~\ref{fig:concept_top_k}, a moderate number of retrieved concepts leads to better performance. 
On MMLongBench, the best result appears around concept Top-$k=10$, while introducing more concepts gradually degrades performance. 
On Qasper, the model remains relatively stable from $k=10$ to $k=40$, but performance drops when $k$ further increases. 
These results suggest that retrieving too few concepts may miss useful graph retrieval directions, whereas retrieving too many concepts can introduce noisy or weakly relevant structures. 
Overall, an appropriate concept Top-$k$ improves QA performance by balancing concept coverage and retrieval focus.
\section{Related Work}
\subsection{RAG on Complex Document}

Retrieval-Augmented Generation (RAG) retrieves external evidence to support answer generation~\cite{lewis2020retrieval,karpukhin2020dense,izacard2021leveraging}. 
Conventional methods such as BM25~\cite{robertson1994some} and VanillaRAG retrieve textual chunks through lexical or dense semantic matching. 
To better handle complex documents, recent methods further incorporate document layout and multimodal content. 
For example, layout-aware retrieval constructs more coherent document units, MM-Vanilla retrieves textual and visual chunks with multimodal embeddings, Tree-Traverse~\cite{zhang2025pageindex} navigates document hierarchies with an LLM, and DocETL~\cite{shankar2024docetl} performs structured document transformations before retrieval. 
Although these methods improve evidence segmentation, they still provide limited modeling of semantic relations among distributed evidence units. 
This limitation becomes critical when reasoning requires evidence aggregation across pages, regions, and modalities.

\subsection{GraphRAG and Multimodal GraphRAG}

GraphRAG~\cite{wang2026omd,niu2026efficientgraph,song2026multimodal} introduces structured relations into retrieval by organizing document knowledge or evidence as graphs. 
RAPTOR~\cite{sarthi2024raptor} builds hierarchical summaries over textual chunks, while GraphRAG~\cite{edge2404local} constructs entity-centric graphs and supports local or global search. 
For multimodal documents, GraphRanker~\cite{gutierrez2024hipporag} applies graph-based ranking to multimodal document graphs, and BookRAG~\cite{wang2025bookrag} combines hierarchical document organization with adaptive retrieval. 
These methods show the benefit of relation-aware evidence modeling. 
However, existing graph-based methods often rely on locally constructed graph elements or dense entity-level traversal. 
As a result, they may suffer from unreliable cross-modal evidence indexing and high retrieval cost. 
In contrast, HVM-GraphRAG introduces a holistic view during graph construction and performs retrieval over a compact concept-level indexing graph.
\section{Conclusion}
This paper proposes Holistic-View Multimodal Graph Retrieval-Augmented Generation (HVM-GraphRAG) on complex document. 
HVM-GraphRAG incorporates a holistic-view into graph construction to address unreliable cross-modal evidence indexing, and uses a compact concept-level graph to guide retrieval for improved efficiency. 
By building reliable links between concept-level graph elements and supporting multimodal chunks, HVM-GraphRAG retrieves grounded evidence while avoiding costly traversal over dense entity-level graphs. 
Experiments on three complex document QA datasets show that HVM-GraphRAG improves both answer performance and online retrieval efficiency over representative RAG baselines. 
These results highlight the importance of reliable evidence indexing and compact graph-guided retrieval for effective complex document QA.

\bibliographystyle{ACM-Reference-Format}
\bibliography{sample-base}

\clearpage
\appendix
\section{Preliminary Study}
\subsection{Definitions}\label{app:definitions}
To facilitate the presentation of our method, we first formalize the core components of our knowledge representation:

\textbf{Concept ($c$) and Entity ($e$)}: A \textit{concept} $c$ denotes a high-level category that abstracts related document entities (e.g., Dataset). 
An \textit{entity} $e$ denotes a concrete instance mentioned in the document (e.g., Qasper). 
Each entity $e$ is assigned to exactly one concept through a concept assignment function $\phi(e)$.

\textbf{Schema ($s$) and Fact ($f$)}:  A \textit{schema} $s=(c_h,r,c_t)$ describes a concept-level relational pattern, where $c_h$ and $c_t$ denote the head and tail concepts, and $r$ denotes the relation between them (e.g., $(\text{Method}, \text{evaluated\_on}, \text{Dataset})$). 
A \textit{fact} $f=(e_h,r,e_t)$ is an entity-level instantiation of a schema, where $e_h$ and $e_t$ denote the head and tail entities connected by the relation $r$ (e.g., (\text{HVM-GraphRAG}, \text{evaluated on}, \text{Qasper})). 
A fact is valid under schema $s$ when $e_h$ and $e_t$ are assigned to the corresponding concepts $c_h$ and $c_t$. 
The modality set is defined as $\Omega=\{\text{text},\text{table},\text{image}\}$.

\textbf{Chunk ($b$)}: A \textit{chunk} $b$ denotes a basic content unit extracted from a complex document. 
Unlike text-only QA settings, a chunk in complex document QA may contain content from different modalities, such as text, tables or images. 
Each chunk is represented as $b=(x,m)$, where $x$ denotes the chunk content and $m$ denotes its modality type (e.g., $m=\text{Table}$).

\subsection{Problem Statement}\label{app:problem_statement}
We study question answering over complex documents. 
Formally, a document $D$ consists of a sequence of $N$ pages, denoted as $D=\{P_i\}_{i=1}^{N}$, where $P_i$ is the $i$-th page. 
These pages contain a set of $M$ multimodal chunks $\mathcal{B}=\{b_j\}_{j=1}^{M}$, where each chunk $b_j$ represents a document unit, such as a text segment, table, or figure. 
A QA system $\mathcal{S}$ maps the document and a query $q$ to an answer $A$:
\[
A = \mathcal{S}(D, q).
\]

Different from QA over short passages, complex document QA often requires evidence distributed across multiple chunks, pages, and modalities. 
This motivates a two-stage GraphRAG process that organizes distributed document evidence offline and retrieves query-relevant evidence online.

\textbf{Offline Graph Structure Construction.}
In the offline stage, the document is transformed into a graph structure for efficient evidence organization. 
Formally, the construction process produces a graph $\mathcal{G}=(\mathcal{V}, \mathcal{E})$ and an evidence index $\mathcal{I}^{*}$:
\[
(\mathcal{G}, \mathcal{I}^{*}) = \text{GraphConstructor}(D),
\]
where $\mathcal{V}$ contains nodes, $\mathcal{E}$ encodes their relations, and $\mathcal{I}^{*}$ maps graph nodes to their supporting multimodal chunks in $\mathcal{B}$. 
The evidence index serves as the bridge between the graph structure and the original document evidence.

\textbf{Online Graph-Guided Retrieval and Answering.}
In the online stage, the system uses the constructed graph and evidence index to answer a user query $q$. 
Instead of retrieving directly from all document chunks, the retriever first localizes query-relevant graph nodes and then expands them to supporting chunks through $\mathcal{I}^{*}$:
\[
\mathcal{B}_q = \text{Retriever}(q, \mathcal{G}, \mathcal{I}^{*}).
\]
The final answer is generated based on the query and the retrieved evidence:
\[
A = \text{LLM}(q, \mathcal{B}_q).
\]

This graph-guided process aims to improve answer accuracy by retrieving semantically connected evidence.

{
\setlength{\dashlinedash}{3pt}
\setlength{\dashlinegap}{2pt}
\setlength{\arrayrulewidth}{0.5pt}

\begin{table}[t]
\centering
\caption{Statistics of the datasets.}
\label{tab:app_dataset_statistics}
\scriptsize
\setlength{\tabcolsep}{3.5pt}
\renewcommand{\arraystretch}{1.15}
\resizebox{0.99\columnwidth}{!}{
\begin{tabular}{cccc}
\toprule
\textbf{Dataset} & \textbf{MMLongBench} & \textbf{M3DocVQA} & \textbf{Qasper} \\
\midrule
Questions   & 669       & 633       & 640 \\
\noalign{\vskip 1.5pt}
\cdashline{1-4}
\noalign{\vskip 1.5pt}
Documents   & 85        & 500       & 192 \\
\noalign{\vskip 1.5pt}
\cdashline{1-4}
\noalign{\vskip 1.5pt}
Avg. Pages  & 42.16     & 8.52      & 10.95 \\
\noalign{\vskip 1.5pt}
\cdashline{1-4}
\noalign{\vskip 1.5pt}
Avg. Images & 25.92     & 3.51      & 3.43 \\
\noalign{\vskip 1.5pt}
\cdashline{1-4}
\noalign{\vskip 1.5pt}
\makecell[c]{Cross-Modal\\Questions (\%)} & 22.73 & 0.00 & 8.31 \\
\noalign{\vskip 1.5pt}
\cdashline{1-4}
\noalign{\vskip 1.5pt}
Tokens      & 2,816,155 & 3,553,774 & 2,265,349 \\
\noalign{\vskip 1.5pt}
\cdashline{1-4}
\noalign{\vskip 1.5pt}
Metrics     & EM, F1    & EM, F1    & Acc, F1 \\
\bottomrule
\end{tabular}
}
\vspace{-4mm}
\end{table}
}

\begin{algorithm*}[t]
\caption{Holistic-View-Guided Graph Construction}
\label{alg:graph_construction}
\begin{algorithmic}[1]
\Require Document $D$; multimodal chunks $\mathcal{B}=\{b_i=(x_i,m_i)\}_{i=1}^{M}$; extraction prompt $\rho_{\mathrm{ext}}^{m_i}$; 
detection prompt $\rho_{\mathrm{det}}$;
resolution prompt $\rho_{\mathrm{res}}$
\Ensure Concept-level indexing graph $\mathcal{G}_{con}=(\mathcal{V}_{con},\mathcal{E}_{con})$; evidence index $
\mathcal I_{\mathrm{con}}(\cdot)$

\Statex \textbf{Stage I: Document Tree Construction}
\State $\mathcal{T}=(\mathcal{N},\mathcal{E}_{\mathcal{T}}) \leftarrow \operatorname{TreeBuilder}(\mathcal{B},D)$
\ForAll{$b_i=(x_i,m_i)\in \mathcal{B}$}
    \State $h_i \leftarrow \operatorname{Path}_{\mathcal{T}}(\operatorname{root}, b_i)$
    \Comment{Obtain document-level structural context}
\EndFor

\Statex \textbf{Stage II: Holistic-View-Guided Multimodal Knowledge Extraction}
\State Initialize holistic view $\mathcal{M}^{0}\leftarrow \emptyset$ and evidence index $\mathcal{I}_{\mathrm{fac}}(\cdot)$
\For{$i=1$ to $M$}
    \State $(\mathcal{C}_i,\mathcal{U}_i,\mathcal{S}_i,\mathcal{F}_i)
    \leftarrow
    \operatorname{Extractor}_{m_i}(h_i, x_i,\rho_{\mathrm{ext}}^{m_i})$
    \Comment{Extract concepts, entities, schemas, and facts}

    \Statex \textbf{Stage III: Conflict Detection and Resolution}
    \State $\mathcal{F}_i^{+}\leftarrow \mathcal{F}_i\cup\mathcal{M}^{i-1}$
    \State Construct $\mathcal{F}_{conf}^{i}$ from $\mathcal{F}_i^{+}$ using matching rules
    \Comment{Same head-relation, same head-tail, or same head}
    \State $\widehat{\mathcal{F}}_{\mathrm{conf}}^i
\leftarrow
\operatorname{Detector}
(
\{g\mid g\in\mathcal{F}_{\mathrm{conf}}^i\},
\rho_{\mathrm{det}}
)$
    \ForAll{$g\in\widehat{\mathcal{F}}_{\mathrm{conf}}^i$}
    \State $\mathcal{Z}_g
\leftarrow
\{e_h,e_t\mid(e_h,r,e_t)\in g\}$
        \State $\mathcal{B}_g
\leftarrow
\bigcup_{e\in\mathcal{Z}_g}
\mathcal I_{\mathrm{fac}}(e)$
        \Comment{Collect supporting chunks}
    \EndFor
    \State $\mathcal{M}^{i},\mathcal{I}_{\mathrm{fac}}^{i}
\leftarrow
\operatorname{Resolver}(\mathcal{M}^{i-1},
\mathcal{I}_{\mathrm{fac}}^{i-1},
\mathcal{F}_i,
\{(g,\mathcal{B}_g)\mid g\in\widehat{\mathcal{F}}_{\mathrm{conf}}^i\},
b_i,\rho_{\mathrm{res}})$
    \Comment{Keep, merge, revise, or remove facts}
\EndFor

\Statex \textbf{Stage IV: Entity-Level Bridging to Concept-Level Indexing}
\State Construct entity-level fact graph from the resolved holistic view $\mathcal{M}$
\State Build concept nodes $\mathcal{V}_{con}$ according to concept assignments
\State Build concept-level edges $\mathcal{E}_{con}$ according to schemas and resolved facts
\ForAll{$z\in \mathcal{V}_{con}$}
    \State $\mathcal{I}_{\mathrm{con}}(z)
\leftarrow
\bigcup_{e\in\phi^{-1}(z)}
\mathcal{I}_{\mathrm{fac}}(e)$
    \Comment{Create direct concept-to-chunk indices}
\EndFor
\State $\mathcal{G}_{con}\leftarrow(\mathcal{V}_{con},\mathcal{E}_{con})$
\State \Return $\mathcal{G}_{con}, \mathcal{I}_{\mathrm{con}}(\cdot)$
\end{algorithmic}
\end{algorithm*}

\section{Experimental Detail} \label{app:exp_det}
\subsection{Dataset Details}\label{app:exp_dataset}
We evaluate \ourmethod on three complex document QA~\cite{zhu2022towards,nagori2025open,sarmah2024hybridrag,scaffidi2025graphrag} datasets: MMLongBench~\cite{ma2024mmlongbench}, M3DocVQA~\cite{cho2024m3docrag}, and Qasp-er~\cite{dasigi2021dataset}. 
These datasets cover different document domains and QA scenarios, including long-form reports, multimodal web-style documents, and scientific papers. 
Following prior work~\cite{wang2025bookrag}, we use the exact processed versions of MMLongBench, M3DocVQA, and Qasper released by the authors, including the same document collection, supplementary questions, and evaluation splits. We introduce no additional data filtering or question augmentation.
The statistics of the three datasets are summarized in Table~\ref{tab:app_dataset_statistics}.

\begin{algorithm*}[t]
\caption{Graph-Guided Holistic Retrieval}
\label{alg:graph_guided_retrieval}
\begin{algorithmic}[1]
\Require Query $q$; concept-level graph $\mathcal{G}_{con}=(\mathcal{V}_{con},\mathcal{E}_{con})$; evidence index $\mathcal{I}_{\mathrm{con}}(\cdot)$; chunk set $\mathcal{B}$; modality set $\Omega=\{\text{text},\text{table},\text{image}\}$; hyperparameters $k_c,\theta,k_r,K,k_b,\{k_m\}_{m\in\Omega}$; answer prompt $\rho_{\mathrm{ans}}$
\Ensure Answer $A$

\Statex \textbf{Stage I: Concept Anchor Retrieval}
\ForAll{$c\in\mathcal{V}_{con}$}
    \State $s(q,c)\leftarrow \operatorname{sim}(\operatorname{Enc}(q),\operatorname{Enc}(c))$
\EndFor
\State $\mathcal{A}_q \leftarrow
\{c \mid (c,s(q,c))\in \operatorname{TopK}_{k_c}(\mathcal{V}_{con};s),\ s(q,c)\geq\theta\}$
\Comment{Select high-confidence concept anchors}

\Statex \textbf{Stage II: Similarity-Based Graph Propagation}
\ForAll{$r$ appearing in $\mathcal{E}_{con}$}
    \State $s(q,r)\leftarrow \operatorname{sim}(\operatorname{Enc}(q),\operatorname{Enc}(r))$
\EndFor
\State $\mathcal{R}_q \leftarrow \operatorname{TopK}_{k_r}(\{r\};s)$
\State $\widehat{\mathcal{R}}_q \leftarrow
\{r\in\mathcal{R}_q \mid
\exists(c_h,r,c_t)\in\mathcal{E}_{con},
\{c_h,c_t\}\cap\mathcal{A}_q\neq\emptyset\}$
\Comment{Keep relations connected to anchors}
\State $\widehat{\mathcal{A}}_q \leftarrow
\mathcal{A}_q \cup
\{c_h,c_t \mid (c_h,r,c_t)\in\mathcal{E}_{con},
r\in\widehat{\mathcal{R}}_q,
\{c_h,c_t\}\cap\mathcal{A}_q\neq\emptyset\}$
\Comment{Expand to related concepts}

\Statex \textbf{Stage III: Graph-Guided Evidence Access}
\State $\mathcal{B}_{q}^{graph}\leftarrow
\bigcup_{z\in\widehat{\mathcal{A}}_q}\mathcal{I}_{\mathrm{con}}(z)$
\Comment{Directly access supporting chunks}
\ForAll{$b\in\mathcal{B}_{q}^{graph}$}
    \State $s(q,b)\leftarrow \operatorname{sim}(\operatorname{Enc}(q),\operatorname{Enc}(\operatorname{Rep}_{m_b}(b)))$
\EndFor
\State $\mathcal{B}_{q}^{K}\leftarrow \operatorname{TopK}_{K}(\mathcal{B}_{q}^{graph};s)$

\Statex \textbf{Stage IV: Conventional RAG Supplementation}
\ForAll{$b\in\mathcal{B}$}
    \State $s(q,b)\leftarrow
    \operatorname{sim}(\operatorname{Enc}(q),\operatorname{Enc}(\operatorname{Rep}_{m_b}(b)))$
\EndFor
\State $\mathcal{B}_{q}^{\mathrm{rag}}\leftarrow
\operatorname{TopK}_{k_b}(\mathcal{B};s)$
\State $\mathcal{B}_{q}^{\mathrm{final}}\leftarrow
\mathcal{B}_{q}^{K}\cup\mathcal{B}_{q}^{\mathrm{rag}}$
\Comment{Merge graph-guided and local evidence}

\Statex \textbf{Stage V: Modality-Aware Evidence Organization}
\ForAll{$m\in\Omega$}
    \State $\mathcal{B}_{q}^{m}\leftarrow
    \{b\in\mathcal{B}_{q}^{\mathrm{final}}\mid m_b=m\}$
    \State $\mathcal{B}_{q}^{m}\leftarrow
    \operatorname{TopK}_{k_m}(\mathcal{B}_{q}^{m};s)$
\EndFor
\State $\mathcal{B}_{q}^{maeo}\leftarrow \bigcup_{m\in\Omega}\mathcal{B}_{q}^{m}$
\State $A \gets \operatorname{Answer}(q,\mathcal{B}_q^{\mathrm{maeo}},\rho_{\mathrm{ans}})$
\State \Return $A$
\end{algorithmic}
\end{algorithm*}

\begin{itemize}
    \item \textbf{MMLongBench:} This is a long-document benchmark designed to evaluate multimodal understanding and QA over lengthy documents. 
    It contains documents from diverse domains, including guidebooks, financial reports, and industry documents. 
    The documents are typically long and contain rich multimodal content, making the dataset suitable for evaluating long-range evidence localization and aggregation.

    \item \textbf{M3DocVQA:} This is a multimodal document question answering dataset constructed from Wikipedia-style HTML documents. 
    It contains heterogeneous document elements such as text, tables, images, captions, and layout structures. 
    The dataset evaluates whether a model can retrieve and QA over multimodal evidence distributed across structured document pages.

    \item \textbf{Qasper:} This is a question answering dataset built from scientific papers. 
    Its questions are information-seeking and often require evidence from different sections of a paper, such as the method, experiment, and result sections. 
    We use Qasper to evaluate QA over scientific papers where supporting evidence may be distributed across long textual contexts and multiple document sections.
\end{itemize}

\subsection{Baselines} \label{exp:baselines}
We compare \ourmethod with representative methods from four RAG categories. \textbf{Conventional RAG} first extracts textual content and divides it into retrieval chunks. This category includes BM25~\cite{robertson1994some} and VanillaRAG. BM25 performs sparse retrieval through lexical matching. VanillaRAG retrieves fixed-size chunks using dense semantic representations. Layout+Vanilla further employs document layout analysis for semantic chunking. \textbf{Graph-Based RAG} extracts textual content and exploits graph structures during retrieval. RAPTOR~\cite{sarthi2024raptor} recursively clusters and summarizes text into a hierarchical structure. GraphRAG~\cite{edge2404local} is evaluated with two search strategies. GraphRAG-Local searches local entity neighborhoods. GraphRAG-Global retrieves information from graph-community summaries. \textbf{Layout-Segmented RAG} partitions documents into structural units through layout analysis. MM-Vanilla retrieves visual and textual units using multimodal embeddings. Tree-Traverse is inspired by PageIndex~\cite{zhang2025pageindex} and uses an LLM to navigate the document tree. DocETL~\cite{shankar2024docetl} provides a declarative framework for complex document processing. \textbf{Multimodal Graph-Structured RAG} includes GraphRanker and BookRAG~\cite{wang2025bookrag}. Graph-Ranker extends HippoRAG~\cite{gutierrez2024hipporag} and applies Personalized PageRank to rank relevant graph nodes. BookRAG organizes multimodal document content into a hierarchical tree and adaptively executes retrieval operators for different queries.
\begin{itemize}
    \item BM25: BM25 is a sparse retrieval method based on lexical matching. 
    It ranks document chunks according to term-frequency and inverse-document-frequency statistics, and retrieves the top-ranked chunks as evidence for answer generation.
    \item Vanilla RAG: Vanilla RAG first divides the document into fixed-size textual chunks and encodes them with a dense retriever. 
Given a query, it retrieves the most semantically similar chunks and feeds them to the generation model.
    \item Layout+Vanilla: Layout+Vanilla extends Vanilla RAG by incorporating document layout information during chunk construction. 
Instead of using only fixed-size text segmentation, it forms more coherent retrieval units based on layout-aware document parsing.
    \item RAPTOR: RAPTOR recursively clusters textual chunks and summarizes them into a hierarchical tree structure. 
During retrieval, it searches over this hierarchy to obtain evidence at different levels of abstraction.
    \item GraphRAG-Local: GraphRAG-Local constructs an entity-centric graph from documents and performs local search around query-relevant entities. 
It retrieves evidence from the neighborhoods of matched entities for answer generation.
    \item GraphRAG-Global: GraphRAG-Global uses graph-community summaries to answer queries that require broader document-level information. 
It retrieves relevant community-level summaries rather than directly searching local entity neighborhoods.
    \item MM-Vanilla: MM-Vanilla extends vanilla dense retrieval to multimodal document chunks. 
It encodes textual and visual units with multimodal representations and retrieves chunks according to query--chunk similarity.
    \item Tree-Traverse: Tree-Traverse organizes document content into a hierarchical document tree and uses an LLM to navigate the tree. 
The retrieved nodes are then used as evidence for final answer generation.
    \item DocETL: DocETL is a declarative document processing framework for complex document analysis. 
It transforms document content into structured intermediate representations before retrieval and answer generation.
    \item GraphRanker: GraphRanker is adapted from HippoRAG, which performs graph-based retrieval over entity-centric knowledge structures. 
To handle complex multimodal documents, GraphRanker extends this idea by constructing a multimodal document graph and applying Personalized PageRank to rank query-relevant graph nodes for evidence retrieval.
    \item BookRAG: BookRAG organizes multimodal document content into a hierarchical structure and adaptively selects retrieval operators for different queries. 
It combines document-structure modeling with multimodal evidence retrieval for complex document question answering.
\end{itemize}

\subsection{Evaluation Metrics}\label{app:Eva_Metric}
In this section, we describe the evaluation metrics used in our experiments. 
Following prior work~\cite{wang2025bookrag}, we evaluate the generated answers with Accuracy, Exact Match (EM), and token-level F1-score. 
Since RAG systems usually generate free-form natural language responses, we first extract and normalize the predicted answer before computing the final scores.

\subsubsection{Answer Extraction and Normalization}
The outputs of RAG-based systems are often natural language sentences rather than short answer spans. 
For example, a model may generate a response such as ``The answer is C'' or ``According to the table, the value is 50.19''. 
Directly comparing such responses with concise ground-truth answers may underestimate model performance. 
Therefore, we follow the standard evaluation setting and apply an answer extraction step before metric calculation.

Let $y_i^{raw}$ denote the raw response generated by the model for the $i$-th query, and let $y_i^{gold}$ denote the corresponding ground-truth answer. 
We use an LLM-based extractor to obtain the concise predicted answer:
\begin{align}
    \hat{y}_i
    =
    \operatorname{LLM}_{extract}(y_i^{raw}, \rho_{ext}),
\end{align}
where $\rho_{ext}$ denotes the extraction instruction, and $\hat{y}_i$ is the extracted answer used for metric computation. 
After extraction, we apply a normalization function $\mathcal{NL}(\cdot)$ to both predictions and ground-truth answers. 
The normalization process includes common text processing operations such as lowercasing, removing punctuation, and eliminating redundant whitespace.

\subsubsection{Accuracy}
Accuracy is used as an inclusion-based soft matching metric. 
A prediction is considered correct if the normalized ground-truth answer is contained in the normalized model response. 
This setting is suitable for free-form LLM outputs, where the generated response may include additional explanatory text around the answer. 
Formally, Accuracy is computed as:
\begin{align}
    \operatorname{Acc}
    =
    \frac{1}{N}
    \sum_{i=1}^{N}
    \mathbb{I}
    \left(
    \mathcal{NL}(y_i^{gold})
    \subseteq
    \mathcal{NL}(y_i^{raw})
    \right),
\end{align}
where $N$ is the number of questions, $\mathbb{I}(\cdot)$ is the indicator function, and $\subseteq$ denotes the substring inclusion relation. 
For Qasper, we report inclusion-based Accuracy as part of our unified evaluation protocol.

\subsubsection{Exact Match}
Exact Match (EM) is a stricter metric than Accuracy. 
It measures whether the normalized extracted answer is exactly the same as the normalized ground-truth answer. 
Given the extracted answer $\hat{y}_i$ and the gold answer $y_i^{gold}$, EM is computed as:
\begin{align}
    \operatorname{EM}
    =
    \frac{1}{N}
    \sum_{i=1}^{N}
    \mathbb{I}
    \left(
    \mathcal{NL}(\hat{y}_i)
    =
    \mathcal{NL}(y_i^{gold})
    \right).
\end{align}
This metric requires character-level equivalence after normalization and is mainly used for datasets with short and unambiguous answers.

\subsubsection{F1-score}
F1-score measures token-level overlap between the extracted answer and the ground-truth answer. 
We first tokenize the normalized extracted answer and the normalized ground truth into token sets $T_{\hat{y}_i}$ and $T_{gold_i}$. 
Precision and recall are then computed as:
\begin{align}
    P_i
    =
    \frac{|T_{\hat{y}_i}\cap T_{gold_i}|}{|T_{\hat{y}_i}|},
    \quad
    R_i
    =
    \frac{|T_{\hat{y}_i}\cap T_{gold_i}|}{|T_{gold_i}|}.
\end{align}
The F1-score for the $i$-th sample is:
\begin{align}
    \operatorname{F1}_i
    =
    \frac{2P_iR_i}{P_i+R_i}.
\end{align}
If both precision and recall are zero, we set $\operatorname{F1}_i=0$. 
The final F1-score is averaged over all questions:
\begin{align}
    \operatorname{F1}
    =
    \frac{1}{N}
    \sum_{i=1}^{N}
    \operatorname{F1}_i.
\end{align}
In our experiments, EM and F1 are reported for MMLongBench and M3DocVQA, while Accuracy and F1 are reported for Qasper following their evaluation protocols.

\subsection{Implementation Details}\label{app:imp_details}
For a fair comparison, all methods are evaluated under the same retrieval and generation setting whenever applicable. 
We use Qwen3-Embedding-0.6B~\cite{zhang2025qwen3} as the embedding model to encode user queries, document chunks, concept nodes, and relations. 
All retrieved evidence is finally provided to the same answer-generation model, Qwen3-8B-AWQ~\cite{yang2025qwen3}, to ensure that performance differences mainly come from retrieval and evidence organization rather than generation capability. 
The same Qwen3-8B-AWQ is also used for answer extraction during evaluation.

\textbf{Document Processing.}
We use \texttt{Mineru}~\cite{wang2024mineru} to parse the input documents and extract layout-aware multimodal chunks. 
Each chunk is associated with its content, modality type, page position, and document structural information. 
The extracted chunks are further organized into a document tree according to title hierarchy, section structure, and reading order. 
This document tree provides the structural context used during graph construction.

\begin{table}[t]
\centering
\caption{Hyperparameter settings of \ourmethod on different datasets.}
\label{app:tab_hyperparameter_settings}
\normalsize
\setlength{\tabcolsep}{4pt}
\renewcommand{\arraystretch}{1.15}
\begin{tabular}{lcccccccc}
\toprule
\textbf{Dataset} & $k_c$ & $\theta$ & $k_r$ & $K$ & $k_b$ & $k_{\text{text}}$ & $k_{\text{table}}$ & $k_{\text{image}}$ \\
\midrule
MMLongBench & 10 & 0.25 & 10 & 10 & 2 & 5 & 3 & 2 \\
M3DocVQA    & 10 & 0.50 & 15 & 10 & 4 & 5 & 3 & 2 \\
Qasper      & 15 & 0.25 & 10 & 10 & 5 & 5 & 3 & 2 \\
\bottomrule
\end{tabular}
\end{table}

\textbf{Graph Construction.}
During offline graph construction, Qwen3-8B-AWQ is used to extract concepts, entities, schemas, and facts from textual chunks. 
For visual chunks, such as figures and images, we use Qwen2.5-VL~\cite{wu2025qwen} as the visual-language extractor. 
The extracted knowledge is incrementally inserted into the holistic view. 
For each newly processed chunk, the system constructs potential conflict groups by comparing newly extracted facts with previously accepted facts. 
Conflict detection and resolution are performed by Qwen3-8B-AWQ. 
The temperature is set to 0 for all LLM and VLM calls to ensure deterministic outputs.

\textbf{Online Retrieval.}
For concept anchor retrieval, we retrieve the top-$k_c$ concept nodes from the concept-level graph and filter low-confidence anchors using the similarity threshold $\theta$. 
We set $k_c=10$ for MMLongBench, $k_c=15$ for Qasper, and $k_c=10$ for M3DocVQA. 
The threshold $\theta$ is set to $0.25$ for MMLongBench and Qasper, and $0.5$ for M3DocVQA. 
For similarity-based graph propagation, we retrieve the top-$k_r$ relations according to their semantic similarity to the query and retain only relations connected to the selected concept anchors. 
We set $k_r=10$ for MMLongBench and Qasper, and $k_r=15$ for M3DocVQA. 
The graph-guided retrieval stage maps selected concept nodes to their supporting chunks through the evidence indexing function. 
The top-$K$ graph-guided chunks are retained with $K=10$ for all datasets. 
For conventional RAG supplementation, we retrieve the top-$k_b$ chunks from the original chunk set, where $k_b=2$ for MMLongBench, $k_b=5$ for Qasper, and $k_b=4$ for M3DocVQA. 
The graph-guided chunks and supplementary chunks are merged and organized by modality. 
For all datasets, we retain $5$ text chunks, $3$ table chunks, and $2$ image chunks in the final modality-aware evidence context.
The detailed hyperparameter settings for each dataset are summarized in Table~\ref{app:tab_hyperparameter_settings}.

\textbf{Baseline Settings.}
For all baseline methods, we follow their original retrieval procedures when available. 
To ensure fairness, all methods use the same embedding model and answer-generation model as \ourmethod. 
For methods requiring top-$k$ retrieval, we set $k=10$ unless otherwise specified. 
For graph-based baselines, including RAPTOR, GraphRAG-Local, GraphRAG-Global, and GraphRanker, we use the same document inputs and generation model as \ourmethod. 
For multimodal and layout-aware baselines, including MM-Vanilla, Tree-Traverse, DocETL, and BookRAG, we use the same parsed document chunks whenever possible.

\textbf{Generation and Reproducibility.}
We do not impose a maximum output length for answer generation. 
The temperature is set to 0 for all generation, extraction, and conflict-resolution calls. 
All experiments are conducted on a machine equipped with an Intel(R) Xeon(R) Gold 5120 CPU and four NVIDIA L40 GPUs with 48 GB memory each. 
All reported results are obtained under the same dataset splits and evaluation protocols described in the main paper.

\begin{table*}[t]
\centering
\caption{Performance comparison with different graph constructors and retrievers.}
\label{app:tab_graph_con}
\renewcommand{\arraystretch}{1.2}
\definecolor{lightblue}{RGB}{232,244,252}
\definecolor{lightpurple}{RGB}{246,239,255}
\definecolor{deltaGreen}{RGB}{0,150,45}

\begin{tabularx}{0.99\textwidth}{
>{\raggedright\arraybackslash}p{0.14\textwidth}
>{\raggedright\arraybackslash}p{0.14\textwidth}
 *{6}{>{\centering\arraybackslash}X}
}
\toprule
\multirow{2}{*}{\textbf{GraphConstructor}}&  \multirow{2}{*}{\textbf{Retriever}} 
& \multicolumn{2}{c}{\textbf{MMLongBench}} 
& \multicolumn{2}{c}{\textbf{M3DocVQA}} 
& \multicolumn{2}{c}{\textbf{Qasper}} \\
\cmidrule(lr){3-4} \cmidrule(lr){5-6} \cmidrule(lr){7-8}
& & \textbf{EM} & \textbf{F1-score} & \textbf{EM} &
\textbf{F1-score} & \textbf{Accuracy} & \textbf{F1-score} \\
\midrule
GraphRAG-Local & GraphRAG-Local
& 7.7  & 8.5  & 23.7 & 25.6 & 35.9 & 39.2  \\

\rowcolor{lightblue}
\ourmethod & GraphRAG-Local
& 9.2 & 10.3 & 25.5 & 27.2 & 37.8 & 40.9  \\

\noalign{\vskip 2pt}
\cdashline{1-8}
\noalign{\vskip 2pt}

GraphRAG-Global & GraphRAG-Global
& 5.3  & 5.6  & 20.2 & 22.0 & 24.0 & 24.1 \\

\rowcolor{lightblue}
\ourmethod & GraphRAG-Global
& 6.0 & 6.7 & 20.8 & 23.1 & 25.2 & 25.6  \\

\noalign{\vskip 2pt}
\cdashline{1-8}
\noalign{\vskip 2pt}

GraphRanker & GraphRanker
& 21.2 & 22.7 & 43.0 & 47.8 & 32.9 & 37.6  \\

\rowcolor{lightblue}
\ourmethod & GraphRanker
& 22.1 &24.0 & 43.4 & 47.9 & 34.3& 39.4 \\

\noalign{\vskip 2pt}
\cdashline{1-8}
\noalign{\vskip 2pt}

BookRAG & BookRAG
& 43.8 & 44.9
& 61.0 & \underline{66.2}
& 55.2 & 61.1  \\

\rowcolor{lightblue}
\ourmethod & BookRAG
&\underline{44.5}& \underline{46.3} & \underline{61.6} & \textbf{66.6} & \underline{56.2} & \underline{62.0} \\

\noalign{\vskip 2pt}
\cdashline{1-8}
\noalign{\vskip 2pt}

\rowcolor{lightpurple}
\ourmethod & \ourmethod
& \textbf{54.1} & \textbf{62.8} & \textbf{63.0} & 66.0& \textbf{66.1}& \textbf{71.5}  \\

\bottomrule
\end{tabularx}
\end{table*}


\subsection{Efficiency Comparison}\label{app:eff_comp}
\begin{figure}[!t]
  \includegraphics[width=1.0\linewidth]{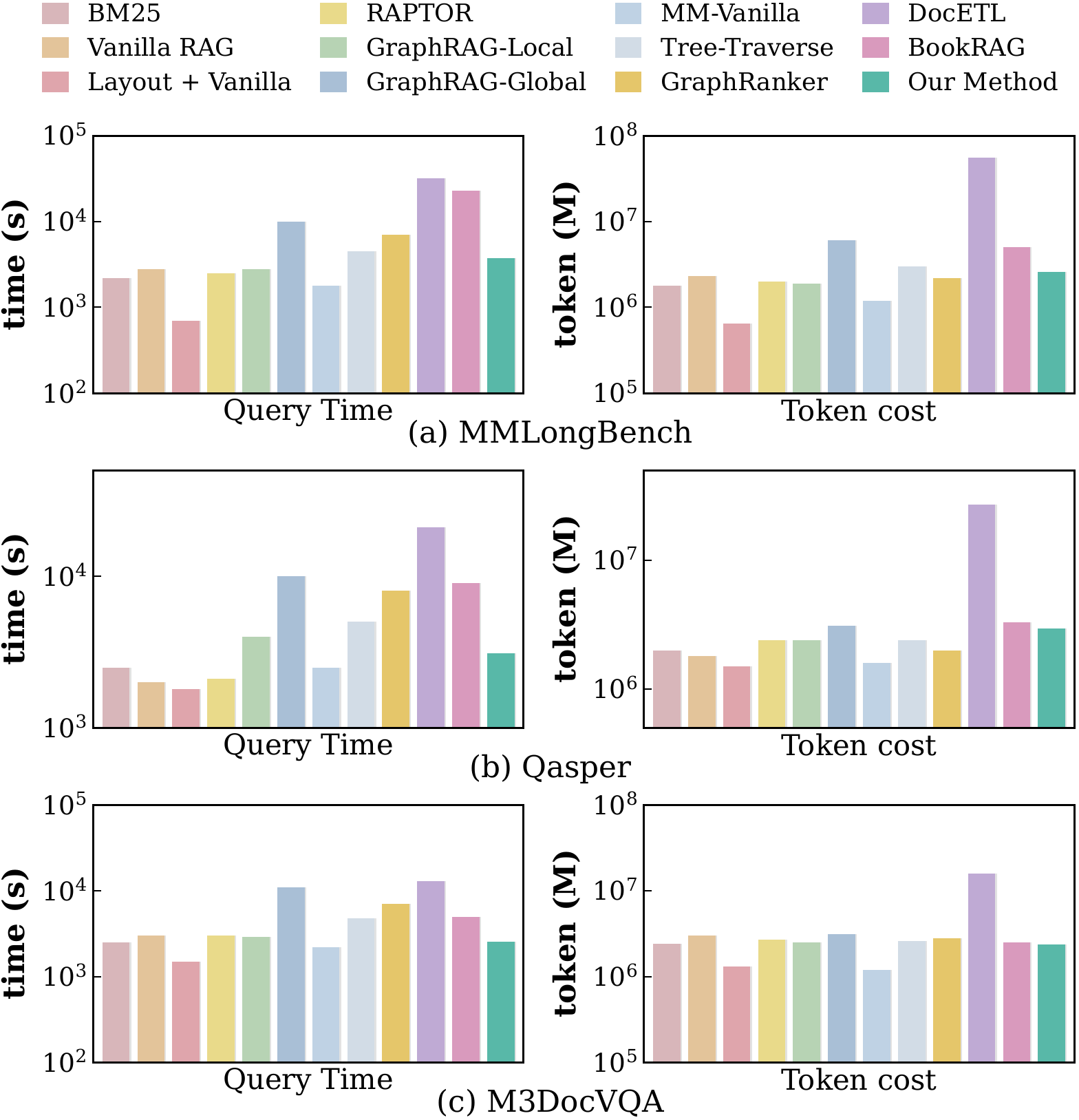}
  \caption{Comparison of query efficiency.}\label{app:fig_efficiency_and_cost}
\end{figure}

Fig.~\ref{app:fig_efficiency_and_cost} reports the total inference time and token consumption of all methods on MMLongBench, Qasper, and M3DocVQA. 
The results provide a more complete view of the efficiency differences among conventional RAG, graph-based RAG, layout-segmented RAG, multimodal GraphRAG, and \ourmethod.

\textbf{Graph-based methods introduce additional online retrieval overhead.}
Compared with conventional RAG methods such as BM25, Vanilla RAG, and Layout+Vanilla, graph-based methods usually require more inference time. 
This trend is especially clear for GraphRAG-Global and GraphRanker across the three datasets. 
The main reason is that these methods need to perform graph-related operations during retrieval, including entity matching, graph traversal, neighborhood expansion, community-level search, or graph ranking. 
Although such operations can improve evidence connectivity, they also increase the computational cost of online retrieval. 
RAPTOR is relatively more efficient than other graph-based methods, but it still requires hierarchical clustering or summary-based retrieval, which makes it less lightweight than simple chunk retrieval.

\textbf{Layout-segmented methods can also be costly.}
Layout-segm-ented methods process document structures more explicitly, but this often leads to more retrieval units or richer intermediate representations. 
As shown in Fig.~\ref{app:fig_efficiency_and_cost}, DocETL incurs particularly high token consumption on all three datasets. 
This is because structured document transformation and declarative processing introduce additional intermediate text and reasoning steps. 
Tree-Traverse also requires more inference time than conventional RAG in most settings, since it relies on LLM-based navigation over document hierarchies. 
These results indicate that layout information is useful for complex document QA, but processing such structures without compact retrieval control may introduce substantial efficiency overhead.

\textbf{BookRAG reduces part of the graph cost but remains expensive.}
BookRAG achieves stronger efficiency than the most expensive graph-based or layout-segmented methods in some cases, but its cost is still higher than conventional RAG. 
This is because BookRAG still relies on hierarchical document traversal and adaptive retrieval operators during online inference. 
Such adaptive retrieval improves flexibility, but it also requires extra LLM calls and increases token consumption. 
Therefore, BookRAG is more efficient than some dense graph traversal methods, but it is still not as lightweight as conventional retrieval.

\textbf{\ourmethod maintains low time and token cost across datasets.}
Across MMLongBench, Qasper, and M3DocVQA, \ourmethod shows query time and token consumption comparable to conventional RAG methods. 
This efficiency comes from two design choices. 
First, \ourmethod performs retrieval over a compact concept-level graph rather than a dense entity-level graph, which reduces the cost of graph matching and traversal. 
Second, once concept nodes are selected, their supporting chunks are directly accessed through the evidence index, avoiding repeated expansion over entity-level nodes. 
Moreover, \ourmethod does not require additional LLM-based query routing or adaptive query analysis during retrieval. 
These designs allow \ourmethod to preserve the benefit of graph-guided evidence modeling while keeping online retrieval cost close to flat retrieval methods.

Overall, the efficiency results support the motivation of \ourmethod. 
Existing graph-based and layout-aware methods often improve evidence organization at the cost of higher inference time or token consumption. 
In contrast, \ourmethod uses a compact concept-level indexing graph to achieve efficient graph-guided retrieval, leading to a better balance between QA performance and computational cost.



\subsection{Generalization of the Graph Constructor}
To further examine the generality of our graph construction strategy, we replace the original graph constructors of several graph-based baselines with the graph construction component of \ourmethod. 
It is worth noting that \ourmethod ultimately builds a concept-level indexing graph, while these baseline methods mainly operate on entity-level graphs. 
For a fair and compatible comparison, we only use the entity-level graph construction part of \ourmethod to replace the original graph constructors of the baselines, while keeping their original retrievers unchanged. 
The results are reported in Table~\ref{app:tab_graph_con}.

\textbf{Our graph constructor consistently improves different graph-based retrievers.}
Replacing the original graph constructor with our construction module improves all baseline retrievers across the three datasets. 
For GraphRAG-Local, the performance increases from 7.7\%/8.5\% to 9.2\%/10.3\% on MMLongBench, from 23.7\%/25.6\% to 25.5\%/27.2\% on M3DocVQA, and from 35.9\%/39.2\% to 37.8\%/40.9\% on Qasper. 
Similar improvements are observed for GraphRAG-Global, GraphRanker, and BookRAG. 
These results suggest that the holistic-view-guided construction process can produce more reliable entity-level graph structures, even when used with retrievers that are not specifically designed for our framework.

\begin{figure*}[t]
\centering
{
{\includegraphics[width=0.32\linewidth]{figures/MMLong_hyparams.pdf}}}
{
{\includegraphics[width=0.32\linewidth]{figures/m3docvqa_hyparams.pdf}}}
{
{\includegraphics[width=0.32\linewidth]{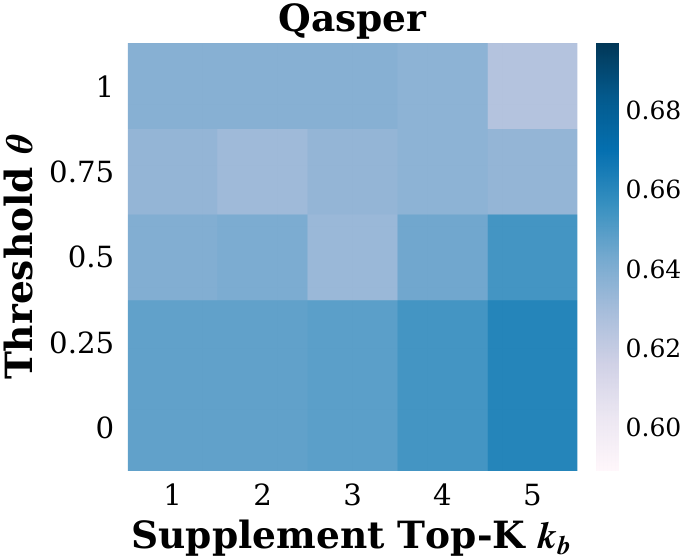}}}
\Description{The edge degree distribution}
\caption{Effects of the similarity threshold $\theta$ and the number of supplementary chunks $k_b$ on MMLongBench, M3DocVQA and Qasper. Darker colors indicate higher performance.}\label{app:fig_threshold_and_supp}
\end{figure*}

\begin{figure*}[t]
\centering
{
{\includegraphics[width=0.32\linewidth]{figures/mmlong_schema_top_k.pdf}}}%
{
{\includegraphics[width=0.32\linewidth]{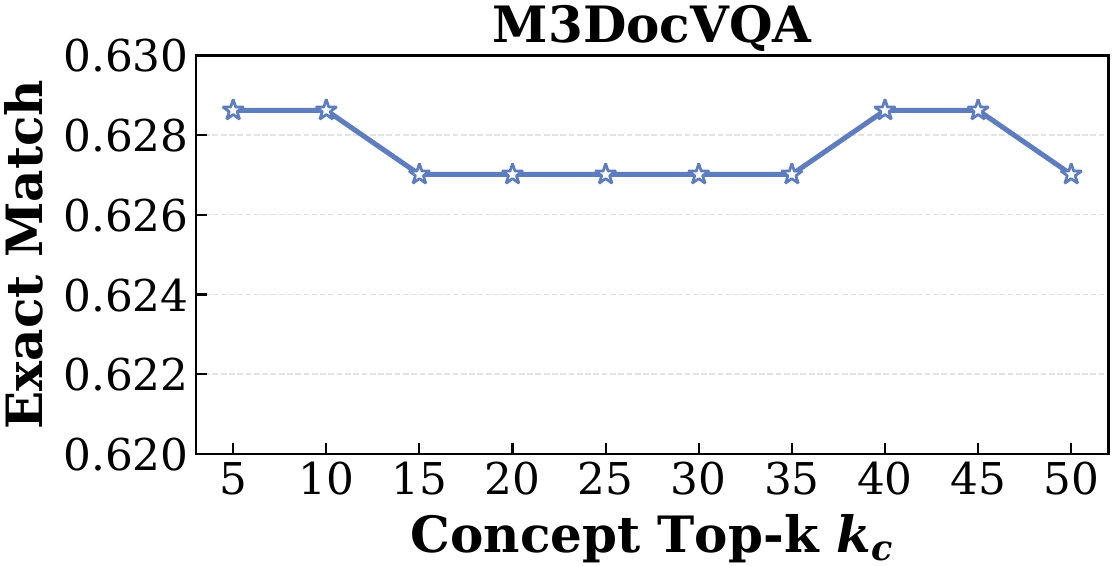}}}%
{
{\includegraphics[width=0.32\linewidth]{figures/qasper_schema_top_k.pdf}}}%
\Description{The edge degree distribution}
\caption{Effects of the number of retrieved concepts $k_c$ on MMLongBench, M3DocVQA and Qasper.}\label{app:fig_concept_top_k}
\end{figure*}

\textbf{The gains are more visible for retrievers that directly depend on local graph structures.}
GraphRAG-Local benefits more clearly from our graph constructor than GraphRAG-Global. 
This is reasonable because GraphRAG-Local retrieves evidence by expanding neighborhoods around matched entities, making it sensitive to the quality of local entity links. 
In contrast, GraphRAG-Global relies on community-level summaries, where local graph improvements may be partially smoothed by aggregation. 
For example, on Qasper, GraphRAG-Local improves by 1.9\% in Accuracy and 1.7\% in F1-score, while GraphRAG-Global improves by 1.2\% in Accuracy and 1.5\% in F1-score.

\textbf{The improvement on stronger multimodal baselines is smaller but still consistent.}
For GraphRanker and BookRAG, replacing the graph constructor also brings stable gains. 
GraphRanker improves from 21.2\%/22.7\% to 22.1\%/24.0\% on MMLongBench and from 32.9\%/37.6\% to 34.3\%/39.4\% on Qasper. 
BookRAG further improves from 43.8\%/44.9\% to 44.5\%/46.3\% on MMLongBench, from 61.0\%/66.2\% to 61.6\%/66.6\% on M3DocVQA, and from 55.2\%/61.1\% to 56.2\%/62.0\% on Qasper. 
These results indicate that our graph construction module can provide cleaner graph evidence even for methods that already consider document structure or graph-based ranking.

\textbf{The full \ourmethod still achieves the best overall performance.}
Although replacing the graph constructor improves existing baselines, the complete \ourmethod remains substantially stronger. 
For example, on MMLongBench, the best constructor-replaced baseline achieves 44.5 EM and 46.3\% F1-score, while the full \ourmethod reaches 54.1\% EM and 62.8\% F1-score. 
On Qasper, the strongest constructor-replaced baseline obtains 56.2\% Accuracy and 62.0\% F1-score, whereas \ourmethod achieves 66.1\% Accuracy and 71.5\% F1-score. 
This gap shows that better graph construction alone is helpful but not sufficient. 
The full advantage of \ourmethod comes from combining holistic-view-guided graph construction, concept-level indexing, graph-guided retrieval, and modality-aware evidence organization.

Overall, this experiment demonstrates that the proposed graph construction strategy has good generality across different graph-based retrieval frameworks. 
At the same time, the superior performance of the full model confirms that reliable graph construction and compact concept-level retrieval are complementary components for complex document QA.

\subsection{Detailed Hyperparameter Analysis}

We provide a more detailed analysis of the retrieval hyperparameters in \ourmethod. 
Fig.~\ref{app:fig_threshold_and_supp} shows the joint effect of the similarity threshold $\theta$ and the number of supplementary chunks $k_b$. 
Fig.~\ref{app:fig_concept_top_k} shows the effect of the number of retrieved concept nodes $k_c$. 
These hyperparameters control different aspects of online retrieval: $\theta$ controls the strictness of concept anchors filtering, $k_b$ controls the amount of complementary evidence from conventional RAG, and $k_c$ controls the coverage of concept-level retrieval.

\begin{figure*}[!t]
  \includegraphics[width=1.0\linewidth]{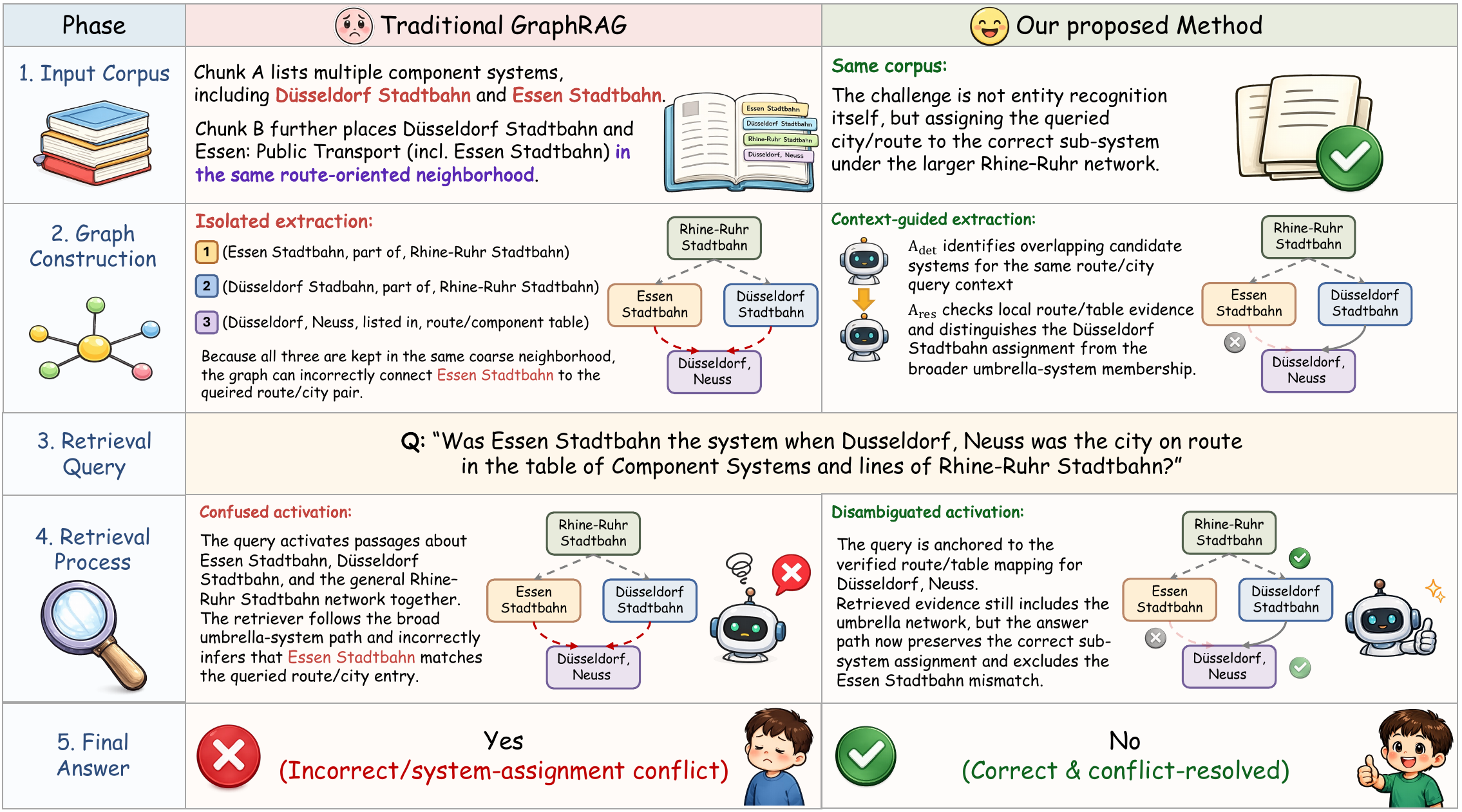}
  \caption{Case Study of Conflict Resolution.}\label{app:cs_conflict}
\end{figure*}

\textbf{Effects of $\theta$.}
The optimal threshold varies across datasets, reflecting their different evidence distributions. 
On MMLongBench, lower thresholds generally lead to better performance. 
The strongest results appear around $\theta=0.25$, especially when $k_b=2$. 
This indicates that long documents benefit from retaining more concept anchors, because useful evidence may be distributed across multiple sections and graph regions. 
A strict threshold may discard weakly matched but still useful concept anchors before graph-guided retrieval begins. 
On Qasper, a similar trend can be observed: lower thresholds such as $\theta=0$ and $\theta=0.25$ often achieve stronger performance, especially when more supplementary chunks are included. 
This suggests that scientific documents also benefit from broader concept anchors coverage, since supporting evidence may be distributed across different sections of a paper. 
In contrast, M3DocVQA performs better under higher thresholds, especially around $\theta=0.5$ to $\theta=1.0$. 
This dataset contains more localized visual question answering cases, where strict concept anchors filtering helps avoid unnecessary graph expansion and keeps retrieval focused on highly matched evidence.

\textbf{Effects of $k_b$.}
The number of supplementary chunks also shows dataset-dependent behavior. 
On MMLongBench, the best performance is achieved with a small number of supplementary chunks, such as $k_b=2$. 
Increasing $k_b$ does not consistently improve performance and may introduce locally similar but less useful chunks in long documents. 
On Qasper, larger $k_b$ values are generally more helpful, with strong performance appearing when $k_b=5$ under low thresholds. 
This suggests that direct query--chunk matching can provide useful complementary evidence for scientific papers. 
On M3DocVQA, performance remains relatively stable as $k_b$ increases, especially under higher thresholds. 
This indicates that the task is less sensitive to supplementary textual retrieval, probably because many questions are grounded in localized visual or page-level evidence. 
Overall, $k_b$ should be viewed as a knob for balancing local evidence supplementation and retrieval noise.

\textbf{Effects of $k_c$.}
Fig.~\ref{app:fig_concept_top_k} shows that a moderate number of retrieved concept nodes usually leads to better performance. 
On MMLongBench, the best result appears around $k_c=10$, and performance generally decreases when more concepts are retrieved. 
This suggests that excessive concept anchors may introduce weakly relevant graph regions in long documents. 
On M3DocVQA, the performance is relatively stable across different $k_c$ values, with slightly stronger results around $k_c=5$, $k_c=10$, $k_c=40$, and $k_c=45$. 
This stability indicates that the dataset is less sensitive to concept-anchor coverage, likely because many questions require localized evidence rather than broad concept-level expansion. 
On Qasper, performance remains stable from $k_c=5$ to $k_c=40$, but drops when $k_c$ increases to $45$ or $50$. 
This suggests that retrieving too many concepts may introduce noisy anchors and weaken retrieval focus.

\textbf{Overall trend.}
The hyperparameter results show that \ourmethod benefits from dataset-specific retrieval settings. 
For long and evidence-distributed documents such as MMLongBench and Qasper, broader concept anchors coverage and a controlled amount of supplementary chunks help improve evidence recall. 
For M3DocVQA, stricter thresholding is more effective because the required evidence is often more localized. 
Across datasets, an overly large concept Top-$k$ is not always beneficial, since excessive concept anchors can introduce irrelevant graph regions. 
These observations support the design of \ourmethod: graph-guided retrieval should balance evidence coverage, retrieval focus, and local evidence supplementation.

\subsection{Case Study}
We provide two qualitative examples to further illustrate how \ourmethod improves complex document QA. 
The first case focuses on conflict detection and resolution during graph construction. 
The second case shows how modality-aware evidence retrieval and organization help the model integrate image, table, and text evidence.

\subsubsection{Case 1: Conflict Resolution in Graph Construction}

Fig.~\ref{app:cs_conflict} illustrates a case where traditional GraphRAG produces an incorrect answer due to unresolved system-assignment conflicts. 
The input document contains multiple component systems under the broader Rhine-Ruhr Stadtbahn network. 
One chunk lists several systems, including D\"usseldorf Stadtbahn and Essen Stadtbahn. 
Another chunk places D\"usseldorf Stadtbahn, Essen-related public transport, and the queried route/city pair within a similar route-oriented context. 
When these chunks are processed independently, traditional GraphRAG keeps all extracted facts in the same coarse neighborhood. 
As a result, the graph incorrectly connects Essen Stadtbahn to the queried D\"usseldorf, Neuss route/city entry.

\begin{figure*}[!t]
  \includegraphics[width=1.0\linewidth]{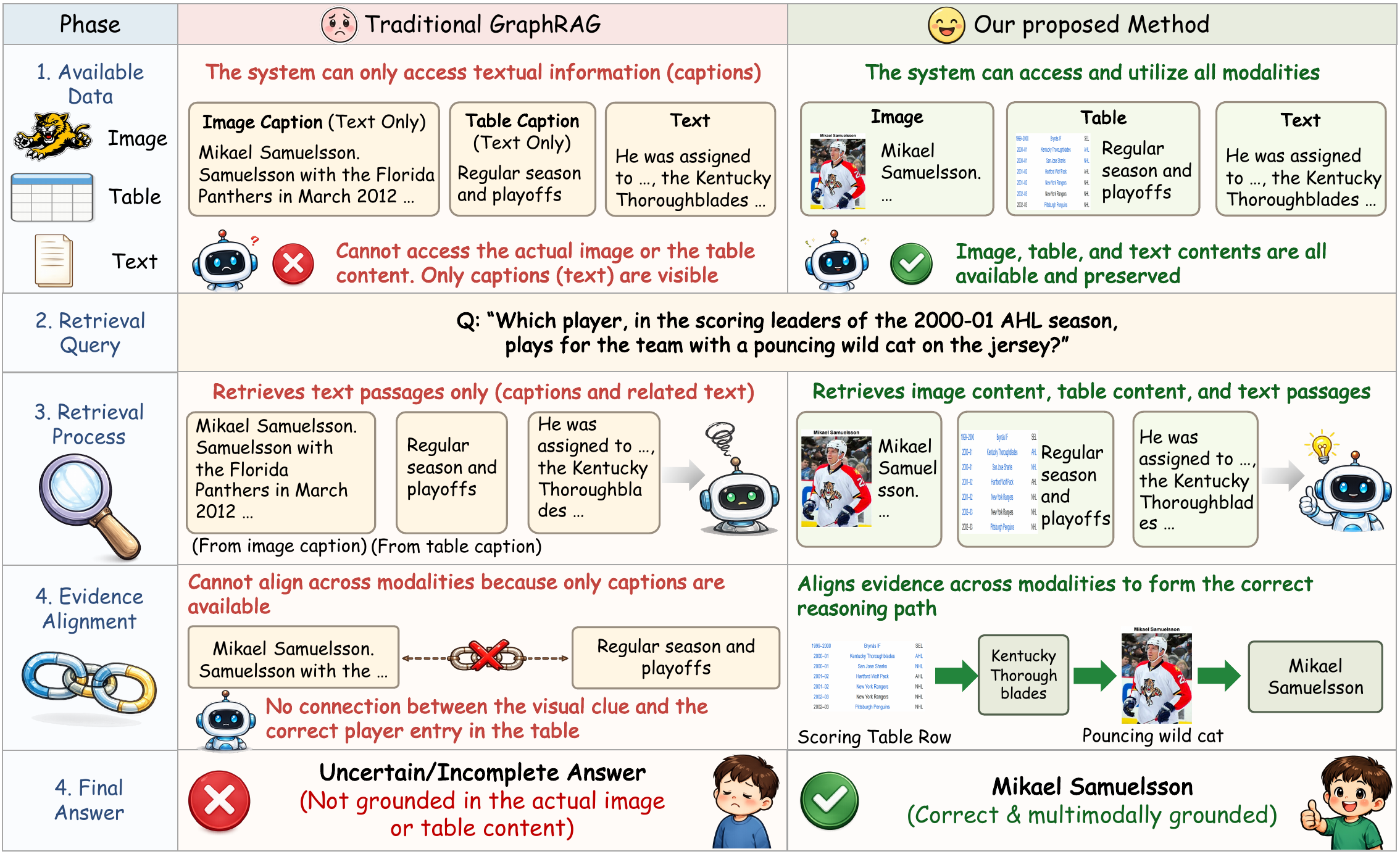}
  \caption{Case Study of Multimodal Evidence Integration.}\label{app:cs_multimodal}
\end{figure*}

During retrieval, the query asks whether Essen Stadtbahn is the system for D\"usseldorf, Neuss in the component-system table of Rhine-Ruhr Stadtbahn. 
Because the graph contains noisy connections among Rhine-Ruhr Stadtbahn, Essen Stadtbahn, D\"usseldorf Stadtbahn, and D\"usseldorf, Neuss, the retriever activates evidence from both the correct and incorrect system assignments. 
The model then follows the broad umbrella-system path and incorrectly answers ``Yes'', treating Essen Stadtbahn as the queried system. 
This shows that graph connectivity alone is insufficient when locally extracted facts introduce ambiguous or conflicting evidence associations.

In contrast, \ourmethod uses the Cross-Modal Holistic View during graph construction to compare newly extracted facts with the existing global fact state. 
For this example, the conflict detector identifies that multiple candidate systems may be associated with the same route or city context. 
The resolver then checks the supporting local evidence and distinguishes the D\"usseldorf Stadtbahn assignment from the broader Rhine-Ruhr umbrella-system membership. 
Consequently, the incorrect Essen Stadtbahn connection is filtered or weakened, while the verified path from D\"usseldorf Stadtbahn to D\"usseldorf, Neuss is preserved. 
During retrieval, the query is anchored to the verified system-route mapping, allowing the model to answer ``No'' correctly. 
This case demonstrates that conflict-aware graph construction can reduce noisy evidence indexing and prevent misleading graph activation during retrieval.

\subsubsection{Case 2: Multimodal Evidence Integration}
Fig.~\ref{app:cs_multimodal} presents a case where the answer requires combining visual, tabular, and textual evidence. 
The query asks which player in the scoring leaders of the 2000--01 AHL season plays for the team with a pouncing wild cat on the jersey. 
Solving this question requires three pieces of evidence: the table identifies the relevant scoring leader and team, the image provides the visual clue of the pouncing wild cat jersey, and the text helps connect the player information with the document context.

Traditional GraphRAG mainly relies on textualized information such as image captions, table captions, and related text passages. 
Although these captions mention Mikael Samuelsson, regular season and playoffs, and related team assignments, they do not expose the actual visual content of the image or the detailed table entries. 
As a result, the retriever cannot reliably align the visual clue in the jersey with the correct player entry in the table. 
The evidence remains fragmented across captions and text snippets, leading to an uncertain or incomplete answer that is not fully supported by the original multimodal content.

\ourmethod retrieves and organizes evidence from multiple modalities. 
The image evidence preserves the visual clue of the jersey, the table evidence provides the scoring-leader row, and the text evidence supplies complementary background information. 
Through graph-guided retrieval, the relevant image, table, and text chunks are jointly selected. 
The modality-aware evidence organization then groups the retrieved chunks by modality, making it easier for the answering model to compare the visual clue with the table entry and textual context. 
This forms a complete reasoning path from the scoring table to the Kentucky Thoroughblades, then to the pouncing wild cat jersey, and finally to Mikael Samuelsson. 
The model therefore produces the correct answer, ``Mikael Samuelsson''. 
This case shows that \ourmethod can better preserve and integrate heterogeneous evidence, especially when the answer depends on information that cannot be recovered from captions or text alone.

Overall, these case studies support the two main design motivations of \ourmethod. 
The conflict-resolution case shows that the holistic view improves the reliability of graph construction and evidence indexing. 
The multimodal QA case shows that graph-guided retrieval and modality-aware evidence organization help the answering model integrate evidence across modalities. 
Together, they explain why \ourmethod achieves more reliable and accurate QA over complex multimodal documents.

\section{Related Work}
\subsection{RAG on Complex Document}
Retrieval-Augmented Generation (RAG)~\cite{fang2026cogniverse,wang2025retrieval,qian2025memorag} has become a widely used paradigm for answering questions over external documents~\cite{robertson1994some,zhang2025pageindex,shankar2024docetl}. 
Early RAG methods usually retrieve evidence from text chunks through sparse or dense matching. 
BM25~\cite{robertson1994some} retrieves relevant chunks based on lexical overlap, while VanillaRAG uses dense semantic representations to match queries with fixed-size document chunks. 
These methods are simple and efficient, but they often ignore the structural organization of complex documents. 
To better preserve document semantics, Layout+Vanilla introduces layout-aware chunking before retrieval, which helps form more coherent retrieval units.

Recent studies~\cite{zhang2025pageindex,shankar2024docetl} further explore layout-segmented and multimodal document retrieval. 
MM-Vanilla extends dense retrieval to both textual and visual document units using multimodal embeddings. 
Tree-Traverse follows the document hierarchy and uses an LLM to navigate structured document trees~\cite{zhang2025pageindex}. 
DocETL~\cite{shankar2024docetl} provides a declarative framework for processing complex document content before retrieval. 
Additionally, recent systems further support iterative visual-document retrieval, multimodal evidence fusion, and heterogeneous document reasoning~\cite{wang2025vidorag,suri2025visdom,yu2025tablerag}.
These methods show that document layout and multimodal information are useful for complex document QA. 
However, they mainly improve how evidence units are segmented or accessed, and still provide limited modeling of semantic relations among distributed evidence units. 
As a result, they may struggle when answering requires multi-hop aggregation across pages, regions, and modalities.

\begin{figure*}[t]
\centering
\begin{promptbox}{Text Entity Extraction}
\scriptsize
\ttfamily

\textbf{-Goal-}

Given a text document that is potentially relevant to this activity and a list of entity types, identify all entities of those types from the text.

\vspace{4pt}
\textbf{-Steps-}

1. Identify all entities. For each identified entity, extract the following information:

\begin{itemize}
    \item entity\_name: Name of the entity, capitalized.
    \item entity\_type: One of the following types: [\{entity\_types\}].
    \item description: A brief, one-sentence description summarizing the entity's role or attributes based \textit{only} on the provided text. If the text offers limited information, keep the description simple, e.g., ``A person mentioned in the text.'' Do not add any external knowledge.
\end{itemize}

2. Return output in English as a single valid JSON object with this exact schema:

\begin{verbatim}
{"entities": [{
      "entity_name": "Entity Name",
      "entity_type": "ENTITY_TYPE",
      "description": "A short description based only on the text."}]}
\end{verbatim}

3. Important output rules:

\begin{itemize}
    \item Output only valid JSON.
    \item Do not output markdown code fences.
    \item Do not output any explanation before or after the JSON.
    \item Do not output tuple-style records.
    \item Do not output relationships.
    \item If no entities are found, return: \{"entities": []\}
\end{itemize}

\vspace{4pt}
\textbf{-Example-}

Entity\_types: [PERSON, ORGANIZATION]

Text:

Yubo Ma $^{{1}}$ , Yixin Cao $^{{2}}$ , YongChing Hong $^{{1}}$ , Aixin Sun $^{{1}}$

Output:

\begin{verbatim}
{"entities": [{
      "entity_name": "Yubo Ma",
      "entity_type": "PERSON",
      "description": "Yubo Ma is a person listed as an author of the document."
    }, ...]}
\end{verbatim}

\vspace{4pt}
\textbf{-Real Data-}

Entity\_types: \{entity\_types\}

Text: \{input\_text\}

Output:

\end{promptbox}

\caption{Prompt for extracting typed entities from textual chunks.}
\label{app_fig_prompt_text_entity_extraction}

\end{figure*}

\begin{figure*}[!h]
\centering
\begin{promptbox}{Table Row Entity Extraction}
\scriptsize
\ttfamily

\textbf{-Goal-}

Act as a precise, row-by-row information extraction engine. Your task is to analyze a batch of \textbf{semi-structured row data strings}. You must \textbf{correlate} each row's data with the provided column\_headers, using the main description for overall context, to extract all relevant entities found strictly within the row data.

\vspace{4pt}
\textbf{-Instructions-}

1. \textbf{Use Context, Don't Extract From It:} First, carefully study the description and column\_headers. Use this information \textbf{strictly as context} to understand the row data's meaning and to assign correct entity types. \textbf{Do not} extract entities that appear only in the description or column\_headers.

2. \textbf{Correlate and Extract from Each Row:} Iterate through each row\_string in the rows\_batch list. For each string:

\begin{itemize}
    \item It represents a single table row, with cells likely separated by a delimiter, e.g., ``|''.
    \item You must mentally map the data cells in the string to their corresponding header in the column\_headers list based on their order.
    \item Extract entities from the row's \textbf{categorical data}, e.g., names of models, methods, or groups like ``CODEX'' or ``SLM + LLM''. You generally \textbf{should NOT} extract entities from purely \textbf{numerical data}, e.g., ``53.8(0.5)''.
    \item \textbf{Minimum Extraction Mandate:} Each row string describes a primary subject. You \textbf{must} aim to extract \textbf{at least one main entity} from each row, representing the subject of that row, e.g., the specific model or method being evaluated.
\end{itemize}

3. \textbf{Consolidate and Format:} Collect all unique entities found across all rows into a single list. Your final response \textbf{must be a single, valid JSON object} with a single root key named entities.

4. \textbf{Strict Extraction Boundaries:} This is the final and most important rule.

\begin{itemize}
    \item All extracted entities \textbf{MUST} originate from the data within the rows\_batch strings.
    \item Therefore, you \textbf{MUST NOT} extract entities that appear only in the description or column\_headers. These are for context only.
    \item You also \textbf{MUST NOT} create an entity for the table itself, e.g., ``Table 1''.
\end{itemize}

\vspace{4pt}
\textbf{-Output JSON Schema-}

\begin{verbatim}
{"entities": [{
      "entity_name": "<String>",
      "entity_type": "<String>",
      "description": "<String>"}]}
\end{verbatim}

\textbf{Field Descriptions}

\begin{itemize}
    \item \textbf{For an Entity object:}
    \item \textbf{entity\_name} String: The primary name of the identified entity, capitalized.
    \item \textbf{entity\_type} String: The category of the entity. It MUST be one of the following types: \{entity\_types\}.
    \item \textbf{description} String: A brief, comprehensive description of the entity's attributes and role in the text.
\end{itemize}

\vspace{4pt}
\textbf{-Example-}

\textbf{Input:}

\begin{verbatim}
{"description": "Table 1: The inference seconds over 500 sentences (run on single V100 GPU).",
  "column_headers": [
    "Dataset (Task)",
    "Roberta",
    "T5",
    "LLaMA",
    "CODEX"],
  "rows_batch": [
    "FewNERD (NER)|2.8|39.4|1135.4|179.4",
    "TACREV (RE)|1.4|45.6|1144.9|151.6",
    "ACE05 (ED)|6.6|62.5|733.4|171.7"]}
\end{verbatim}

\textbf{Output:}

\begin{verbatim}
{"entities": [{
      "entity_name": "FewNERD (NER)",
      "entity_type": "DATASET_OR_CORPUS",
      "description": "A Named Entity Recognition dataset used as a benchmark to measure model inference speeds."
    },...]}
\end{verbatim}

\vspace{4pt}
\textbf{-Task Execution-}

Analyze the following input data using the allowed entity types provided. Your response must be the complete JSON object and nothing else.

\vspace{4pt}
\textbf{-Allowed Entity Types-}

\{entity\_types\}

\vspace{4pt}
\textbf{-Input to Process-}

\{input\_json\}

\vspace{4pt}
\textbf{-JSON Output-}

\end{promptbox}
\caption{Prompt for extracting typed entities from table rows using table descriptions and column headers as context.}
\label{app:fig_prompt_table_row_entity_extraction}
\end{figure*}

\subsection{GraphRAG and Multimodal GraphRAG}
Graph-based RAG methods~\cite{zhang2025graph,yu2026graphrag} introduce graph structures to organize document knowledge and support relation-aware retrieval. 
RAPTOR~\cite{sarthi2024raptor} recursively clusters and summarizes textual chunks into a hierarchical structure, enabling retrieval at different levels of abstraction. 
GraphRAG~\cite{edge2404local} constructs entity-centric graphs from documents and supports both local and global search strategies. 
GraphRAG-Local retrieves information from neighborhoods around query-relevant entities, while GraphRAG-Global answers queries using graph-community summaries. 
Compared with flat chunk retrieval, these methods can better capture semantic connections among evidence units. 
However, their retrieval process often depends on entity matching, graph traversal, or community-level summarization, which can introduce additional cost and may be less effective for multimodal document evidence.

Multimodal GraphRAG~\cite{gutierrez2024hipporag,wang2025bookrag} further extends graph-based retrieval to complex documents containing text, tables, figures, and other modalities. 
GraphRanker~\cite{gutierrez2024hipporag} adapts graph-based ranking to multimodal document graphs and applies Personalized PageRank to identify relevant graph nodes. 
BookRAG~\cite{wang2025bookrag} organizes multimodal document content with hierarchical structures and adaptively selects retrieval operations for different queries. 
These methods demonstrate the potential of structured retrieval for complex document QA. 
Nevertheless, existing multimodal graph-based methods usually focus on building or traversing graph structures, while paying less attention to the reliability of links between graph elements and their supporting multimodal chunks. 
In contrast, our work emphasizes holistic-view-guided graph construction and reliable concept-level evidence indexing, so that graph-guided retrieval can be both accurate and efficient.

\section{Prompt}
This section presents the prompt templates used in \ourmethod. 
These prompts support two major stages of our framework: multimodal knowledge extraction and holistic-view-guided conflict handling. 
For knowledge extraction, we design modality-specific prompts for textual chunks, table rows, and image chunks, so that each type of document content can be processed under suitable extraction constraints. 
For conflict handling, we design prompts for detecting and resolving inconsistent facts during graph construction. 
All prompts require structured JSON outputs, which enables automatic parsing and reduces uncontrolled free-form generations.

\subsection{Multimodal Knowledge Extraction}

\textbf{Text Entity Extraction Prompt.}
Fig.~\ref{app_fig_prompt_text_entity_extraction} shows the prompt used to extract typed entities from textual chunks. 
Given a text chunk and a predefined list of entity types, the prompt asks the LLM to identify all entities that belong to the allowed types and to generate a short description for each extracted entity. 
The description is required to be based only on the provided text, preventing the model from introducing external knowledge. 
The prompt also enforces a strict JSON schema with the key \texttt{entities}, which contains the entity name, entity type, and entity description. 
If no valid entity is found, the model must return an empty entity list. 
This prompt is mainly used for text-like chunks, including paragraphs and textual descriptions extracted from complex documents.

\textbf{Table Row Entity Extraction Prompt.}
Fig.~\ref{app:fig_prompt_table_row_entity_extraction} presents the prompt for extracting entities from table rows. 
Different from ordinary text extraction, table understanding requires the model to interpret each row together with its column headers and table description. 
Therefore, the prompt provides the table description and column headers as contextual information, while explicitly restricting entity extraction to the row content. 
This design prevents the model from incorrectly extracting entities that only appear in headers or captions. 
The prompt further asks the model to map row cells to their corresponding columns and to extract categorical entities such as methods, datasets, systems, or organizations, while avoiding pure numerical values. 
The output follows the same JSON format as the text entity prompt, making table entities compatible with the later graph construction process.

\textbf{Image Entity Extraction Prompt.}
Fig.~\ref{app:fig_prompt_image_entity_extraction} shows the prompt used for extracting entities from visual chunks. 
For image-based content, the prompt is given to a VLM and requires the model to jointly analyze the image and its accompanying textual description. 
The prompt first asks the model to include the image itself as an entity, which preserves a direct index from the visual chunk to the graph. 
It then instructs the model to identify visual objects, labels, annotations, titles, and other meaningful elements appearing in the image. 
When the accompanying description contains useful context, the model may also use it to clarify the role of visual entities. 
This prompt allows \ourmethod to preserve visual information that cannot be fully recovered from captions alone.


\textbf{Relation Extraction Prompt.}
Fig.~\ref{app:fig_prompt_relation_extraction} gives the prompt for extracting relations among the entities identified from a chunk. 
Given the chunk content and the extracted entities, the prompt asks the LLM to identify high-confidence directional relations. 
Each relation is represented as a structured triple with a head entity, relation name, tail entity, and their corresponding entity types. 
The prompt requires relation names to be normalized using lowercase words with underscores, and it instructs the model to avoid weak, speculative, or unsupported relations. 
This step converts extracted entities into entity-level facts, which are later used to construct the entity-level fact graph and establish direct indices from concept nodes to their supporting chunks.

\subsection{Conflict Detection and Resolution}

\textbf{Conflict Detection Prompt.}
Fig.~\ref{app:fig_prompt_conflict_detection} presents the prompt for detecting conflicts among candidate fact groups. 
During graph construction, newly extracted facts may overlap or conflict with facts already accepted in the Cross-Modal Holistic View. 
The prompt asks the LLM to examine each candidate group and determine whether the triples truly conflict. 
It considers three conflict types: mutual conflict, temporal conflict, and granularity conflict. 
Mutual conflict captures mutually exclusive facts, temporal conflict captures time-dependent facts with incompatible or missing time scopes, and granularity conflict captures facts expressed at different levels of specificity. 
The prompt requires the model to return whether a conflict exists, the conflicting triple pairs, the conflict type, and a short explanation. 
This design avoids treating all similar facts as conflicts and helps distinguish actual contradictions from compatible facts.

\textbf{Conflict Resolution Prompt.}
Fig.~\ref{app:fig_prompt_conflict_resolution} shows the prompt used to resolve detected conflicts. 
Given conflicting triples and their supporting chunks, the prompt asks the LLM to decide whether each triple should be kept, discarded, or modified. 
For mutual conflicts, the resolver selects the fact better supported by the evidence. 
For temporal conflicts, it may modify the relation with time information when such evidence is available. 
For granularity conflicts, it may preserve compatible facts by adding scope information to the relation. 
The output contains resolved triples, unresolved conflicts, and a brief summary of the resolution process. 
By using supporting chunks during resolution, this prompt helps \ourmethod reduce noisy or conflicting graph updates and maintain a more reliable evidence index.

\begin{figure*}[t]
\centering
\begin{promptbox}{Image Entity Extraction}
\scriptsize
\ttfamily

\textbf{-Goal-}

Act as an expert AI system for visual and semantic analysis. Your primary task is to \textbf{comprehensively} analyze a given image and an accompanying textual description. You must identify \textbf{all possible} relevant entities, formatting the output as a single, valid JSON object that adheres to the provided Pydantic schema.

\vspace{4pt}
\textbf{-Steps-}

1. \textbf{Analyze Full Context:} Carefully examine both the provided image and the textual description. The description often provides crucial context, names, or other details that are not visually obvious.

2. \textbf{Identify Entities:}

\begin{itemize}
    \item \textbf{Crucial First Step: The Image Itself.} You \textbf{MUST} identify the entire image as the very first entity. Its entity\_type \textbf{must} be IMAGE and it \textbf{must} be the first object in the final entities list. \textbf{This is a non-negotiable rule.} If the description provides a title or figure number, e.g., ``Figure 2'', use that as the entity\_name; otherwise, use the first few words of the description as the entity name. If the description is empty, use ``The Image'' as the entity name.

    \item \textbf{From Visual Objects:} Identify distinct physical objects, people, animals, and general locations shown in the image.

    \item \textbf{From Text Within the Image High Priority:} Pay attention to any text inside the image, such as labels, titles, annotations, or data points in diagrams and flowcharts. \textbf{This text is a critical source of entities.} You \textbf{MUST} treat every distinct label, title, or significant term as a candidate for an entity.

    \item \textbf{From the Description:} Extract any additional relevant entities mentioned in the text that might not be visible or clearly identifiable in the image.

    \item \textbf{Principle of Comprehensiveness:} When in doubt, it is better to extract a potential entity than to omit it. \textbf{Be thorough and aim for maximum detail.}

    \item \textbf{Create Entity Objects:} For each unique entity found, create a JSON object following the ExtractEntity structure.
\end{itemize}

3. \textbf{Construct the Final Output:} Combine all identified entities into a single JSON object with the root key ``entities'', ensuring it strictly follows the specified output format.

\vspace{4pt}
\textbf{-Output Format-}

\textbf{1. General Instruction}

Your response MUST be a single, valid JSON object with the root key ``entities''.

\vspace{4pt}
\textbf{2. JSON Structure}

\begin{verbatim}
{
  "entities": [
    {
      "entity_name": "<String>",
      "entity_type": "<String>",
      "description": "<String>"
    }
  ]
}
\end{verbatim}

\textbf{3. Field Descriptions}

\begin{itemize}
    \item \textbf{entity\_name} String: The name of the entity.
    \item \textbf{entity\_type} String: The category of the entity. It MUST be one of the following types: \{entity\_types\}.
    \item \textbf{description} String: A brief, comprehensive summary of the entity's role and attributes.
\end{itemize}

\vspace{4pt}
\textbf{-Example-}

\textbf{Conceptual Input:}

\begin{itemize}
    \item \textbf{Image:} A simple diagram showing three boxes. ``User PC'' has an arrow pointing to ``Web Server'', which has an arrow pointing to ``Database''.
    \item \textbf{Description:} ``Figure 1: A diagram of a basic three-tier web application architecture.''
\end{itemize}

\textbf{Correct Output:}

\begin{verbatim}
{
  "entities": [
    {
      "entity_name": "Figure 1",
      "entity_type": "IMAGE",
      "description": "A diagram illustrating a basic three-tier web application architecture, showing the data flow between components."
    },
    {
      "entity_name": "User PC",
      "entity_type": "SYSTEM_COMPONENT",
      "description": "The client-tier component in the architecture diagram, representing the end-user's machine."
    },
    ...
  ]
}
\end{verbatim}

\vspace{4pt}
\textbf{-Task-}

Now, analyze the provided image and its description. Generate the JSON object according to all the instructions above.

\vspace{4pt}
\textbf{-Allowed Entity Types-}

\{entity\_types\}

\vspace{4pt}
\textbf{-Image description-}

\{image\_description\}

\vspace{4pt}
\textbf{-JSON Output-}

\end{promptbox}
\caption{Prompt for extracting typed entities from images and accompanying textual descriptions.}
\label{app:fig_prompt_image_entity_extraction}
\end{figure*}

\begin{figure*}[t]
\centering
\begin{relationpromptbox}{Relation Extraction}
\footnotesize
\ttfamily

You are an expert in knowledge graph construction and relation extraction.
Your task is to extract relations between given entities from the provided text.

\vspace{4pt}
\textbf{-Instructions-}

1. Analyze the text carefully and identify meaningful relations between the provided entities.

2. Only extract relations that are explicitly stated or strongly implied in the text.

3. Use lowercase and underscores for relation names, e.g., ``works\_for'', ``located\_in''.

4. Ensure relations are directional: subject\_entity, relation, object\_entity.

5. Do not invent relations that are not supported by the text.

6. Return only high-confidence relations. Skip weak, generic, repetitive, or speculative relations.

7. Do not repeat the same relation triple. If the same head, relation, tail appears multiple times, return it only once.

8. Prefer specific factual relations such as affiliation, citation, authorship, part\_of, type\_of, evaluates, uses, improves, inspired\_by, or related\_to only when the text strongly supports them.

9. Do not create placeholder entities, inferred aliases, or broad summary relations that are not directly grounded in the text.

10. Return at most 20 distinct relations, ordered from most explicit/high-confidence to least.

\vspace{4pt}
\textbf{-Output Format-}

Return exactly one valid JSON object with this structure:

\begin{verbatim}
{
  "relations": [
    {
      "head": "the subject entity (exactly as provided)",
      "relation": "the relation type",
      "tail": "the object entity (exactly as provided)",
      "head_type": "the entity type/label of the head entity (exactly as provided)",
      "tail_type": "the entity type/label of the tail entity (exactly as provided)"
    }
  ]
}
\end{verbatim}

The ``relations'' value must always be a JSON array.

If multiple valid relations exist, include all of them in the array.

If no valid relations can be extracted, return:

\begin{verbatim}
{"relations": []}
\end{verbatim}

Before returning, deduplicate the array so that each relation triple appears only once.

\vspace{4pt}
\textbf{-Example input-}

\textbf{Text:}

Apple Inc. was founded by Steve Jobs in 1976.

\vspace{2pt}
\textbf{Entities to consider:}

\begin{itemize}
    \item Apple Inc. (ORG)
    \item Steve Jobs (PERSON)
    \item 1976 (DATE)
\end{itemize}

\textbf{-Example output-}

\begin{verbatim}
{
  "relations": [
    {
      "head": "Apple Inc.",
      "relation": "founded_in",
      "tail": "1976",
      "head_type": "ORG",
      "tail_type": "DATE"
    },
    {
      "head": "Steve Jobs",
      "relation": "founded",
      "tail": "Apple Inc.",
      "head_type": "PERSON",
      "tail_type": "ORG"
    }
  ]
}
\end{verbatim}

Do not return markdown, explanations, or any keys other than ``relations''.

\end{relationpromptbox}
\caption{Prompt for extracting high-confidence relations between given entities from textual chunks.}
\label{app:fig_prompt_relation_extraction}
\end{figure*}

\begin{figure*}[t]
\centering
\begin{conflictpromptbox}{Conflict Detection}
\scriptsize
\ttfamily

\textbf{System:}

You are an expert fact checker. Given one candidate conflict group of triples, determine whether any triples in the group truly conflict with each other.

Your task:

Analyze the triples within the group together, compare them pairwise where useful, and detect real conflicts. Classify conflicts into three types:

\begin{itemize}
    \item mutual conflict: mutual exclusivity / one-to-one relations
    \item temporal conflict: time-dependent facts; conflicts arise when time scopes overlap or are missing
    \item granularity conflict: different levels of specificity; may be compatible via containment/hypernymy
\end{itemize}

\textbf{-Definitions and rules-}

\textbf{1) Mutual conflict, type = ``mutual''}

A mutual conflict happens when:

\begin{itemize}
    \item Same subject and predicate, but different objects, AND the predicate is one-to-one / mutually exclusive.

    Example: (X, birthplace, Shanghai) vs (X, birthplace, Beijing)

    \item Or cyclic/contradictory relational structure that cannot both be true under common-sense constraints.

    Example: (A, father, B) vs (B, father, A)
\end{itemize}

\textbf{2) Temporal conflict, type = ``temporal''}

A temporal conflict happens when:

\begin{itemize}
    \item The predicate describes a role/state that can change over time and is typically unique at a given moment, e.g., president/CEO/champion/current location.

    \item If both triples claim different objects for the same subject-predicate:
    \begin{itemize}
        \item If explicit time scopes exist and overlap $\rightarrow$ hard temporal conflict.
        \item If time scopes exist and do NOT overlap $\rightarrow$ not a conflict.
        \item If time scopes are missing but the predicate is time-variant and moment-unique $\rightarrow$ suspected temporal conflict. Ask for time ranges; do NOT assert a hard conflict without time info.
    \end{itemize}
\end{itemize}

\textbf{3) Granularity conflict, type = ``granularity''}

A granularity conflict happens when:

\begin{itemize}
    \item Triples differ due to specificity/abstraction level.

    Example: (X, birthplace, Shanghai) vs (X, birthplace, China)

    \item If one object is a parent/superset/contains the other, i.e., hypernym/meronym/administrative containment, then it is usually compatible $\rightarrow$ classify as ``granularity''.

    \item If objects are incompatible, cannot contain each other, and cannot both be true $\rightarrow$ logical conflict.
\end{itemize}

\textbf{-Important instructions-}

\begin{itemize}
    \item Only judge conflicts among the triples inside the provided group.
    \item Use the group type and possible conflict reason as hints, not as proof of conflict.
    \item Do not mark a conflict unless it is clearly supported by the triples.
    \item If there is no real conflict in the group, return has\_conflict=false.
\end{itemize}

Output MUST be a valid JSON object following the required schema.

\vspace{4pt}
\textbf{User:}

Analyze the following candidate conflict group.

\begin{verbatim}
Group ID:
${group_id}

Group Type:
${group_type}

Group Scope:
${group_scope}

Possible Conflict Reason:
${group_reason}

Triples in Group:
${group_triples}
\end{verbatim}

Output a JSON object with the following structure:

\begin{verbatim}
{
  "has_conflict": true/false,
  "conflicts": [
    {
      "triple1_id": "id1",
      "triple1": ["head", "relation", "tail"],
      "triple2_id": "id2",
      "triple2": ["head", "relation", "tail"],
      "conflict_type": "mutual|temporal|granularity",
      "conflict_reason": "brief explanation of why these triples conflict"
    }
  ],
  "conflicting_triple_ids": ["id1", "id2", ...]
}
\end{verbatim}

Analyze the whole group. If conflicts exist, list all conflicting triple pairs.

If has\_conflict is false, return empty arrays for conflicts and conflicting\_triple\_ids.

JSON payload:

\end{conflictpromptbox}
\caption{Prompt for detecting conflicts among candidate groups of extracted triples.}
\label{app:fig_prompt_conflict_detection}
\end{figure*}

\begin{figure*}[t]
\centering
\begin{resolvepromptbox}{Conflict Resolution}
\scriptsize
\ttfamily

\textbf{System:}

You are an expert knowledge graph curator. Given a set of conflicting triples and their source passages, your task is to resolve the conflicts and produce corrected triples.

\vspace{4pt}
\textbf{-Conflict Resolution Strategies-}

\textbf{1. Mutual Conflict, type = ``mutual'':}

\begin{itemize}
    \item These are contradictory claims about the same entity, e.g., same subject-predicate but different objects.
    \item Resolution: Analyze the source passages to determine which triple is more accurate.
    \item Keep only the CORRECT triple, discard the incorrect one(s).
    \item If both seem equally valid based on context, prefer the one with more specific/credible source.
\end{itemize}

\textbf{2. Temporal Conflict, type = ``temporal'':}

\begin{itemize}
    \item These are time-dependent facts where time scopes overlap or are missing.
    \item Resolution: Add time information to the relation to distinguish the facts.
    \item Modify the predicate to include time context, e.g., ``was president of [2000-2005]'' vs ``was president of [2005-2010]''.
    \item If time info is not in sources, note it as ``temporal\_conflict\_unresolved''.
\end{itemize}

\textbf{3. Granularity Conflict, type = ``granularity'':}

\begin{itemize}
    \item These are facts at different levels of specificity, e.g., ``born in Shanghai'' vs ``born in China''.
    \item Resolution: Add granularity description to the relation to clarify the scope.
    \item Modify the predicate to include granularity context, e.g., ``was born in [city: Shanghai]'' vs ``was born in [country: China]''.
    \item Both can be kept if they are compatible, i.e., containment relationship.
\end{itemize}

Output MUST be a valid JSON object following the required schema.

\vspace{4pt}
\textbf{User:}

Resolve the following conflicting triples using their source passages.

\begin{verbatim}
Conflicting Triples and Their Sources:
${conflicting_triples_with_sources}
\end{verbatim}

Output a JSON object with the following structure:

\begin{verbatim}
{
  "resolved_triples": [
    {
      "original_triple": ["head", "relation", "tail"],
      "triple_id": "fact_id",
      "conflict_type": "mutual|temporal|granularity",
      "resolution": "kept|discarded|modified",
      "resolved_triple": ["head", "modified_relation", "tail"] or null if discarded,
      "reason": "explanation of why this resolution was chosen"
    }
  ],
  "unresolved_conflicts": [
    {
      "triple_ids": ["id1", "id2"],
      "reason": "reason why conflict could not be resolved"
    }
  ],
  "summary": "brief summary of how conflicts were resolved"
}
\end{verbatim}

For each conflicting triple:

\begin{itemize}
    \item If resolution is ``kept'': Keep the triple as is, because it is correct.
    \item If resolution is ``discarded'': The triple is incorrect, set resolved\_triple to null.
    \item If resolution is ``modified'': Provide the modified triple with time/granularity information in the relation.
\end{itemize}

JSON payload:

\end{resolvepromptbox}
\caption{Prompt for resolving conflicting triples using their source passages.}
\label{app:fig_prompt_conflict_resolution}
\end{figure*}
\end{document}